# Data-Driven Topology Optimization for Multiscale Biomimetic Spinodal Design

Shiguang Deng [a], Doksoo Lee [b], Aaditya Chandrasekhar [b], Stefan Knapik [b], Liwei Wang [c], Horacio D. Espinosa [b], Wei Chen [b] *

[a] Department of Mechanical Engineering, University of Kansas, Lawrence, KS, USA

[b] Department of Mechanical Engineering, Northwestern University, Evanston, IL, USA

[c] Department of Mechanical Engineering, Carnegie Mellon University, Pittsburgh, PA, USA

## Abstract

Spinodoid architected materials have drawn significant attention due to their unique nature in stochasticity, aperiodicity, and bi-continuity. Compared to classic periodic truss-, beam- and plate-based lattice architectures, spinodoids are insensitive to manufacturing defects, scalable for high throughput production, functionally graded by tunable local properties, and material failure resistant due to low-curvature morphology. However, the design of spinodoids is often hindered by the curse of dimensionality with extremely large design space of spinodoid types, material density, orientation, continuity, and anisotropy. From a design optimization perspective, while genetic algorithms are often beyond the reach of computing capacity, gradient-based topology optimization is challenged by the intricate mathematical derivation of gradient fields with respect to various spinodoid parameters. To address such challenges, we propose a data-driven multiscale topology optimization framework. Our framework reformulates the design variables of spinodoid materials as the parameters of neural networks, enabling automated computation of topological gradients. Additionally, it incorporates a Gaussian Process surrogate for spinodoid constitutive models, eliminating the need for repeated computational homogenization and enhancing the scalability of multiscale topology optimization. Compared to 'black-box' deep learning approaches, the proposed framework provides clear physical insights into material distribution. It explicitly reveals why anisotropic spinodoids with tailored orientations are favored in certain regions, while isotropic spinodoids are more suitable elsewhere. This interpretability helps to bridge the gap between data-driven design with mechanistic understanding. To this end, we test our design framework on several numerical experiments. We find our multiscale spinodoid designs with controllable anisotropy achieve better performance than single-scale isotropic counterparts, with clear physics interpretations.

## 1. Introduction

Existing truss-, beam- and plate-based architected materials, i.e., mechanical metamaterials, have proven great success [1] due to their simple geometries, predictable behaviors, and ease of mass production; however, they often suffer from catastrophic failures caused by stress concentrations at stress-raisers, material interfaces, and defects, resulting in low strength, short fatigue life, or buckling under extreme loading. On the contrary, biomimetic spinodoids generally resemble the grand diverse nature of biological materials with unique features, such as stochasticity, aperiodicity, anisotropy, and irregularity, as in Figure 1(a). Spinodoids are considered superior to conventional cellular materials due to their high resistance to stress concentrations and reduced sensitivity to manufacturing imperfections [2], [3]. However, the design of such spinodoid material systems is hindered by prohibitively large design spaces that often require impractically high sampling efforts to surrogate underlying functions, aka the curse of dimensionality. Compared to gradient-free methods, e.g., genetic algorithms, gradient-based Topology Optimization (TO) is generally more efficient for architected material design. But when applied to spinodoid designs, it faces two challenges: ($i$) gradient development often involves intricate and error-prone mathematical derivations and ($ii$) TO repeatedly needs costly computation of effective constitutive properties.

To address the challenges, we propose a data-driven design framework of spinodoid architected material systems by systematically integrating several state-of-the-art design innovations: ($i$) a Neural-Networks-based multiscale TO with automatic differentiation to account for various design gradients for both categorical variables, e.g., types of spinodoid, and continuous design parameters, such as material density, orientation, continuity, and anisotropy, ($ii$) a spectral density function-based microstructure reconstruction to efficiently generate an ensemble of statistically equivalent stochastic spinodoid microstructures from an interpretable low-dimensional space where a robust interface interpolation scheme enhances the bi-continuity across microstructural boundaries, and ($iii$) a Gaussian Process surrogate of spinodoid constitutive models that bypasses repeated expensive homogenization within TO loops. In this process, we generate a 2D spinodoids data repository containing multiple types of spinodoid microstructures with varying volume fractions, orientations, and anisotropy levels. We demonstrate our method on several volume-constrained compliance minimization experiments, as exemplified in Figure 1(b); however, we note that our design framework is extensible to other types of objectives and constraints, accommodating general spinodoid designs with various design parameters.

*Corresponding author.
Email address: weichen@northwestern.edu

The rest of our paper is organized as follows. In Section 2, we review existing works on spinodoid architected materials and multiscale TO, and discuss the research gaps that we aim to address. In Section 3, we provide an overview of our approach where technical details are provided. In Section 4, we evaluate the performance of our proposed method in several numerical experiments, and we conclude the paper in Section 5.

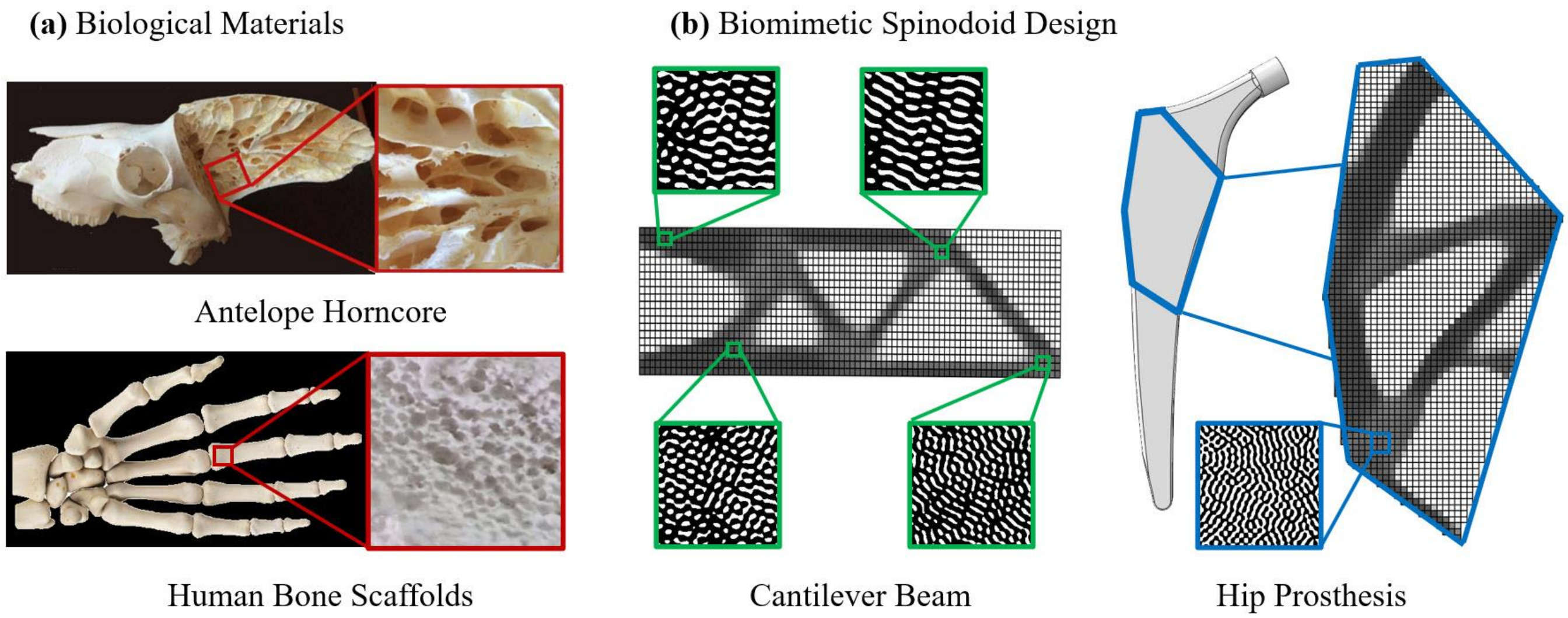

**Figure 1 Biological materials and biomimetic spinodoid architected materials: (a)** Biological materials exhibit very distinct morphology than truss-, beam- or plate-based engineered materials. They are often resulted from long-term evolution to fulfill various biological functions, such as sustaining mechanical loads, being lightweight, and ensuring biocompatibility [4] where figures are adapted from [5] and [6] licensed under CC BY 4.0, **(b)** Spinodoids, resembling spinodal patterns from spinodal decomposition, often manifest similar material morphology as biological materials. The aim of this work is to achieve a unified design framework for spinodoid material-structure systems by integrating local spinodoids design with structural optimization.

## 2. Literature Review

Spinodoids are a unique class of materials that emerge from the self-assembly process driven by phase separation, involving the spontaneous evolution of multiphase mixtures [7], [8], [9]. Through self-assembly, spinodoids can be mass-produced to create centimeter-scale samples with nanoscale features, while avoiding the defects that typically arise in high-resolution additive manufacturing processes [10]. Such multiscale spinodoids manifest superior mechanical properties than many conventional periodic lattice materials. For instance, researchers have found that aperiodic spinodoid morphologies can exhibit stretching-dominant behaviors [4], [7], offering superior mechanical properties, including enhanced stiffness, strength, and toughness, across a wide range of material densities [7], while maintaining insensitivity to local morphological variations [3]. Additionally, its high surface-to-volume ratio could also be advantageous in multi-functional devices, like sound and energy absorptions [11], [12], and biological functionalities such as fluid transport, bioactivity and osseointegration [4]. Even though the spinodoids could potentially revolutionize the current architected material systems, their designs are often intractable, as it is nearly impossible to gain a full exploration of the vast anisotropic design space due to its lack of periodicity and symmetry [2]. One of the earliest attempts is to use spinodal decomposition simulations to capture the time-dependent Cahn-Hilliard phase separation process for target morphology design [13]. While such process simulation-based design approaches are typically too expensive, more practical spinodoid design often involves machine learning and/or TO.

Machine learning has drawn significant attention in material designs to bypass expensive process simulations. Some major efforts include purely data-driven approaches for surrogating material constitutive modeling [14], [15] and physics-constrained neural networks (NNs), which directly solve governing equations and integrate physical laws as soft constraints within machine learning architectures [16], [17], [18], [19]. Inverse designs of spinodoids for target properties have been solved via tandem NNs [2], generative modeling [20], [21], semi-supervised learning [21], and variational autoencoders [22], [23]. However, machine learning models often work as a black box, relying on statistics and the quality of data, and may lack physics-consistent design insights and interpretation.

TO originates as a structure optimization technique [24] and has rapidly advanced to include multiscale modeling to account for material-structure system design with hierarchical microstructures [25], [26], [27], [28], [29]. Unlike single-scale TO, multiscale TO requires careful consideration of the connectivity between locally oriented microstructures to ensure structural integrity and performance. This often requires simultaneously optimizing the geometries at all spatial scales [30], [31], or decomposition methods to map locally oriented microstructures in a smoothly varying field [32], [33], [34]. In a recent study, a multiscale TO approach was developed for 3D spinodoid design, where four types of 3D spinodoids were created and each spinodoid microstructure was characterized by pre-defined wave vectors, porosity densities, and orientations [4]. The connectivity between different spinodoids in [4] was ensured by projecting phase fields of spinodoids to a unified function where microstructure interfaces were smoothed via interface interpolations. Albeit successful, one of its limitations [4] is that its sensitivity analysis was performed analytically using elemental quantities of interest, which could be non-trivial when TOs involve a large number of multiple different types of design variables. Another pioneer work of spinodal design was reported in [35] where a two-scale topology optimization was developed for the design of macroscopic bodies with optimal elastic responses. Particularly, [35] used Gaussian Random Fields, inspired from phase field analyses to simulate phase separations in spinodal decomposition process, to generate spinodal microstructures. [35] directly manipulates the rigid body rotations of microstructures during topology optimization to control their local directions, and it utilized Deep Neural Networks to surrogate elastic properties by spinodal geometric descriptors where the gradient of elastic moduli can be computed with respect to geometric descriptors via automatic differentiation.

Compared to machine learning methods, TO is a mechanics-based optimization scheme with mechanistic governing equations and interpretable design insights. However, multiscale TOs are challenged by intricate and error-prone mathematical derivations of design gradients, and repetitive costly homogenization of microstructural properties. To address such challenges, our work is to develop a data-driven TO framework for spinodoid architected materials with automatic differentiation. The advantages of our work are three-fold: ($i$) we propose a novel microstructure reconstruction method to generate multiple types of stochastic spinodoid microstructures; ($ii$) we create a Gaussian Process data-driven surrogates of microstructural effective properties to avoid repeated computational homogenization in TO loops; and ($iii$) we develop a Neural Networks-based TO that can automatically perform sensitivity analyses with respect to various design variables for both isotropic and anisotropic spinodoids in multiscale material-structure designs.

## 3. Proposed Framework

Our proposed design framework is illustrated in Figure 2. It comprises several key components, including a generic NN that maps spatial coordinates of microstructures to their spinodal design variables, and reformulates the design variables as network's weight parameters, a Gaussian Process regression model, trained offline using a data repository generated from stochastic spinodoid microstructure reconstructions, for the online prediction of constitutive properties, a SIMP-based penalization scheme that penalizes elastic properties based on material densities and spinodoid types, and an automatic differentiation module that automates the computation of topological gradients with respect to NNs' weights. After topology optimization, the interface smoothness between neighboring microstructures is enhanced by boundary interpolation and amplifying functions to improve manufacturability.

In the following sections, we first discuss technical details on how to reconstruct stochastic spinodoid microstructures in Section 3.1. We then proceed to the enforcement of smooth boundary transitions and the calculation of effective elastic properties in Sections 3.2 and 3.3, respectively. Then, we elaborate on the dataset generation of the spinodoids repository and the surrogate training in Section 3.4. We explain the details of multiscale neural TO in Section 3.5, and provide detailed execution steps of the proposed design framework at the end of this section.

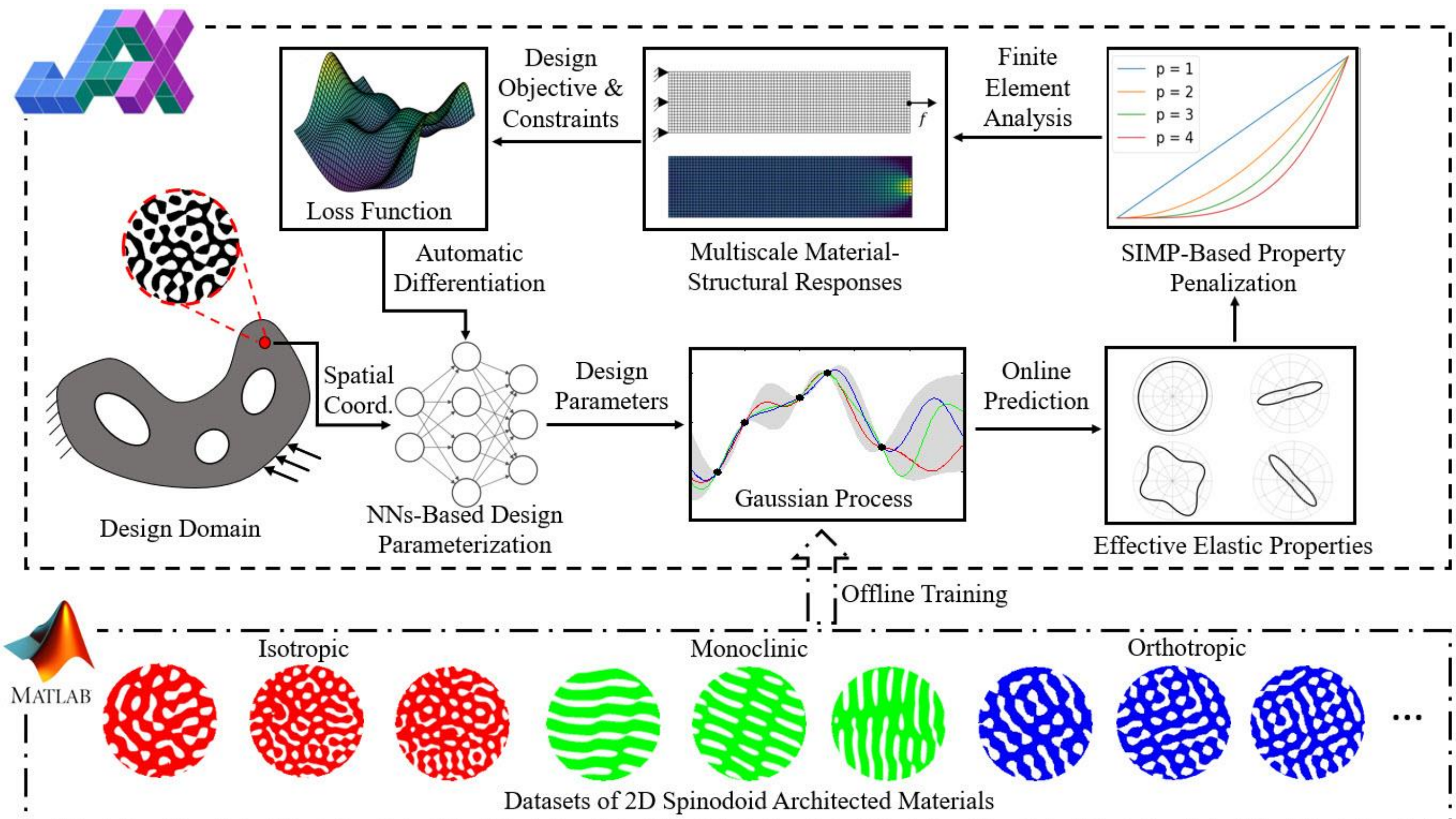

**Figure 2 Our proposed data-driven design framework for the multiscale topology optimization with biomimetic spinodoid architected materials.**

### 3.1. Microstructure Reconstruction of Stochastic Spinodoids

Our microstructure reconstruction is based on the Spectral Density Function (SDF), which can provide a low-dimensional physically meaningful frequency-based description of quasi-random material systems [36], [37], [38]. Our SDF-based spinodoid microstructure reconstruction is inspired by the Linear Time Invariant (LTI) model from control theory [39]. It states that the output signal can be deduced from a known impulse and arbitrary input signal [40] as:

$$S_o(\xi) = |H|^2 S_i(\xi), \tag{1}$$

where $S_i$ and $S_o$ are the SDFs of input and output signals with the frequency of $\xi$, and $H$ is the Fourier Transform of an impulse response. In addition, the SDF can be computed as the squared magnitude of the Fourier Transformation of the signal $\varphi$ [40] as:

$$S(\xi) = |F[\varphi]|^2, \tag{2}$$

where $F[\cdot]$ is the Fourier Transform operator. Since the phase field of a microstructure can be considered as a 2D signal $\varphi$, the microstructure can be reconstructed via SDFs from Equation (1), as follows:

$$S_R(\boldsymbol{k}) = S_T(\boldsymbol{k}) \bullet S_W(\boldsymbol{k}), \tag{3}$$

where $S_R$, $S_T$, and $S_W$ are the SDFs of reconstructed, target, and white noise signals (we can consider the signals as the phase fields of microstructures), and $\boldsymbol{k}$ is the spatial frequency vector. That is, $\boldsymbol{k}$ is the vector version of the scalar frequency of $\xi$ in Equation (1). The operator "•" represents point-wise multiplication, and the reconstructed, target and white noise microstructures must have the same spatial resolutions. By comparing Equation (1) with Equation (3), it is clear the target phase field $S_T$ resembles the impulse $H$, while the reconstructed and white noise phase fields ($S_R$ and $S_W$) correspond to the input and output signals ($S_i$ and $S_o$), respectively.

We note that as the SDF is the squared magnitude of the Fourier Transform, the phase field of the reconstructed microstructure can be computed as:

$$\begin{aligned} \varphi_R &= \left| F^{-1}\left[ \sqrt{S_T(\boldsymbol{k})} \bullet F[\varphi_W] \right] \right| \\ &= \left| F^{-1}\left[ |F[\varphi_T]| \bullet F[\varphi_W] \right] \right|, \end{aligned} \tag{4}$$

where the phase field of the reconstructed microstructure $\varphi_R$ can be computed by applying Fourier Transformations on the phase fields of the target microstructure $\varphi_T$ and white noise microstructure $\varphi_W$, respectively. We note that the use of white noise microstructure $\varphi_W$ introduces stochasticity, since there exist multiple different realizations of $\varphi_W$ depending on the probabilistic distribution of the white noise. In addition, the white noise microstructure $\varphi_W$ can be considered as a white noise image containing all frequencies in an equal measure. Therefore, the reconstruction process of Equation (4) can be interpreted as that the target signal $\varphi_T$ performs an convolution operation on the white noise signal $\varphi_W$ to filter out all frequencies except for the target frequencies in $\varphi_T$. For this reason, the phase field of the reconstructed microstructure $\varphi_R$ is considered as a random realization of $\varphi_T$.

To obtain the binary image of the reconstructed microstructure, we continue to use a level-set function to cut its phase field $\varphi_R$ as:

$$\Lambda(\boldsymbol{X}) = \begin{cases} 1, & \text{if } \varphi(\boldsymbol{X}) \le \varphi_{cut}(\rho) \\ 0, & \text{otherwise,} \end{cases} \tag{5}$$

where $\Lambda(\boldsymbol{X})$ is the binary image of the reconstructed spinodoid microstructure containing only one of the two values, i.e., either 1 when the phase field $\varphi$ is smaller or equal to the cutting plane $\varphi_{cut}$ (corresponding to the desired microstructural volume fraction $\rho_m$), or 0 when the $\varphi$ is larger than the cutting plane value, at each location of its spatial domain $\boldsymbol{X}$. We note that if $\varphi_T$ is periodic, the reconstructed microstructure $\Lambda(\boldsymbol{X})$ also preserves periodicity.

By using Equations (4) and (5), we can reconstruct isotropic spinodoids by using the two parameters: the microstructure material density $\rho_m$ and the spatial frequency vector $\boldsymbol{k}$ as in Figure 3(a). We then proceed to develop two additional types of anisotropic spinodoids: monoclinic spinodoids with unidirectional anisotropy and orthotropic spinodoids with bidirectional anisotropy. To build a monoclinic microstructure, we need two parameters to control the directional anisotropy of underlying phase fields: a frequency rotational angle $\gamma$ and an anisotropic index $\alpha^{\mathrm{mon}}$ as:

$$\alpha^{\mathrm{mon}} = \sin\left(\theta^{\mathrm{mon}}\right), \tag{6}$$

where the anisotropic index $\alpha^{\mathrm{mon}}$ is the sine function of a polar angle $\theta^{\mathrm{mon}}$, representing the percentage of missing frequencies from the isotropic state in the phase field, as shown in Figure 3(b). The rotational angle $\gamma$ guides directional anisotropy by measuring the skewness from the vertical direction. The $\alpha^{\mathrm{mon}}$ in Equation (6), therefore, can take the value in the range of [0, 1] where 0 represents full isotropy and 1 indicates an absolute anisotropic state (whose morphology may deviate from the spinodoids, therefore should be avoided, as discussed in Sections 3.5).

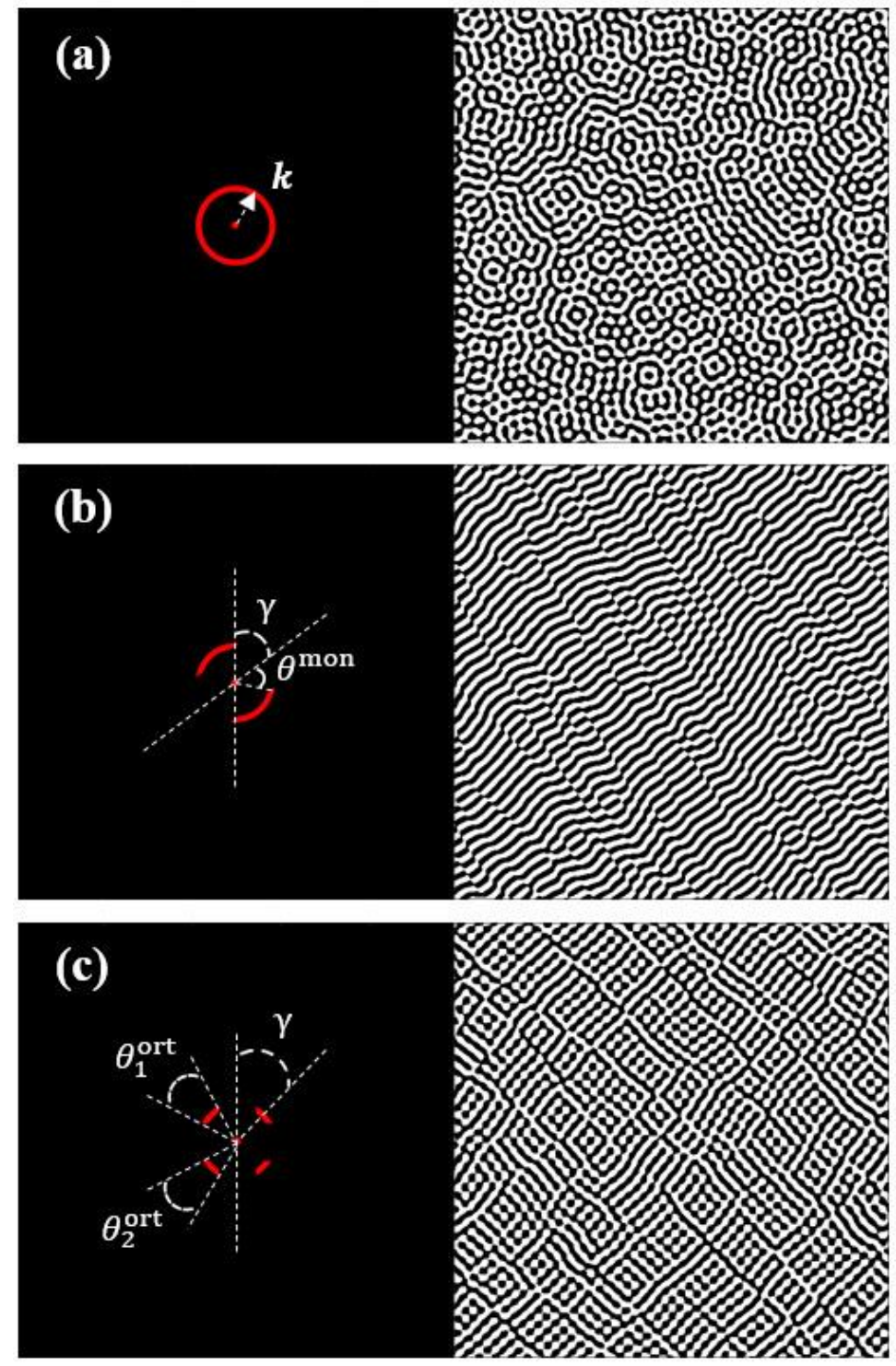

**Figure 3 Illustration of SDFs, spinodoids design parameters and their corresponding microstructures: (a)** In addition to material volume fraction, the isotropic spinodoids are characterized by spatial frequency $\boldsymbol{k}$ (illustrated as the radius vector), **(b)** Monoclinic spinodoids are characterized by the rotational angle $\gamma$ and polar angle $\theta^{\mathrm{mon}}$ parameters, and **(c)** Orthotropic spinodoids are characterized by the rotational angle $\gamma$ and two orthogonal polar angle $\theta_1^{\mathrm{ort}}$ and $\theta_2^{\mathrm{ort}}$ parameters, respectively. For both monoclinic and orthotropic spinodoids, 0 anisotropic index indicates isotropy, while 1 represents absolute anisotropy. Red rings indicate the target frequencies on the SDF domains.

To develop an orthotropic spinodoid microstructure, we need to replace the polar angle $\theta^{\mathrm{mon}}$ of the monoclinic spinodoids with two separate polar angles $\theta_1^{\mathrm{ort}}$ and $\theta_2^{\mathrm{ort}}$, and define the anisotropic indices $\alpha_1^{\mathrm{ort}}$ and $\alpha_2^{\mathrm{ort}}$ for orthotropic spinodoids as:

$$\alpha_1^{\mathrm{ort}} = \cos\left(\theta_1^{\mathrm{ort}}\right), \qquad \alpha_2^{\mathrm{ort}} = \cos\left(\theta_2^{\mathrm{ort}}\right), \tag{7}$$

where $\theta_1^{\mathrm{ort}}$ and $\theta_2^{\mathrm{ort}}$ are two orthogonally oriented polar angles measuring the percentage of the nonzero frequencies in the phase field, as shown in Figure 3(c). Similarly, the values of $\alpha_1^{\mathrm{ort}}$ and $\alpha_2^{\mathrm{ort}}$ are also in the ranges of $[0,1]$. When $[\alpha_1^{\mathrm{ort}}, \alpha_2^{\mathrm{ort}}] = [1,1]$, it results in absolute anisotropy in two orthogonal directions. $[\alpha_1^{\mathrm{ort}}, \alpha_2^{\mathrm{ort}}] = [0,0]$, it returns to an isotropic state. But when one of the anisotropic indices takes the value of 0, the orthotropic microstructure is reduced to a monoclinic state.

We demonstrate the effects of model parameters on the three types of spinodoids as shown in Figure 4. The first row of Figure 4 shows three realizations of isotropic spinodoids. Observe that Figure 4(a-1) and (a-2) have the same density $\rho_m$ but different frequency $\boldsymbol{k}$ (shown as the radius of the ring of the SDF field, which is color coded as red), and Figure 4(a-2) and (a-3) have different $\rho_m$ but the same $\boldsymbol{k}$. It is obvious that increasing $\boldsymbol{k}$ results in a smaller spinodoid feature size, while increasing $\rho_m$ increases material percentage, but not change local morphology. We use the same frequency $\boldsymbol{k}$ for the demonstration of monoclinic and orthotropic spinodoids in the second and third rows of Figure 4. By comparing Figure 4(b-1) and (b-2), it is obvious that increasing the anisotropic index $\alpha^{\text{mon}}$ results in a stronger unidirectional microstructure, while changing rotational angle $\gamma$ in Figure 4(b-2) and (b-3) clearly changes material distribution directions. Additionally, as we increase $\alpha_1^{\text{ort}}$ and $\alpha_2^{\text{ort}}$ from Figure 4(c-1) to (c-2), we can clearly see clear bidirectional material distributions. Similarly, setting the rotational angle $\gamma$ can help guide the material distribution directions as in the monoclinic case.

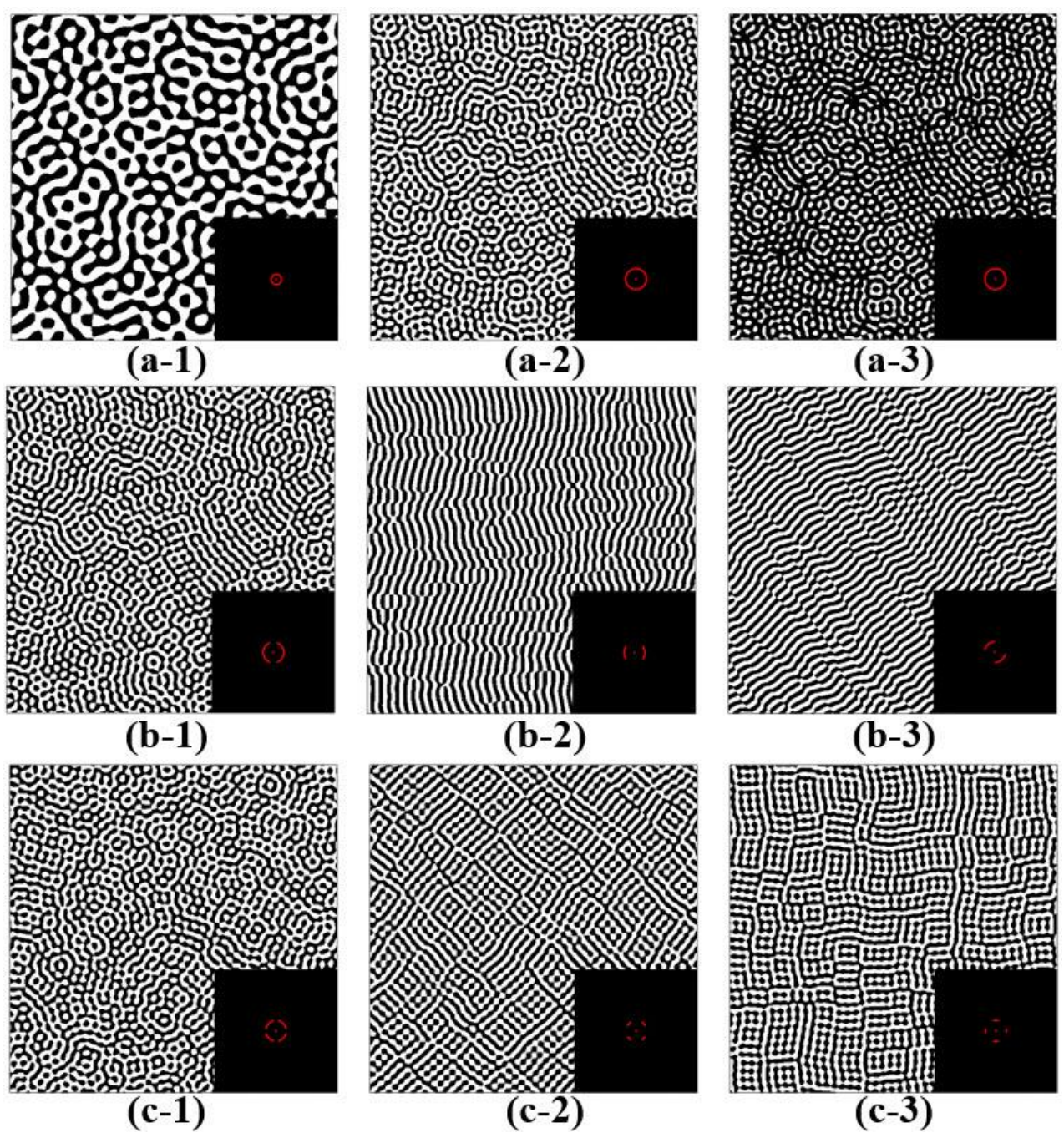

**Figure 4 Different morphologies of the three types of spinodoids with different parameters: (a)** Isotropic spinodoids have mostly identical material distributions in all directions where black pixels indicate materials while white ones represent voids, **(b)** Monoclinic spinodoids exhibit strong unidirectional dependency, and **(c)** Orthotropic spinodoids manifest material orientations along two orthogonal directions.

### 3.2. Boundary Connection Across Spinodoid Microstructures

As functionally graded materials (FGMs), spinodoids often have spatially varying morphologies to obtain locally different properties. But when we combine multiple reconstructed spinodoid microstructures from Section 3.1, their boundary interfaces are often not well connected due to the periodicity of each microstructure, as illustrated in Figure 5. The ill-connected interface would cause manufacturing difficulty and stress singularity in extreme mechanics scenarios.

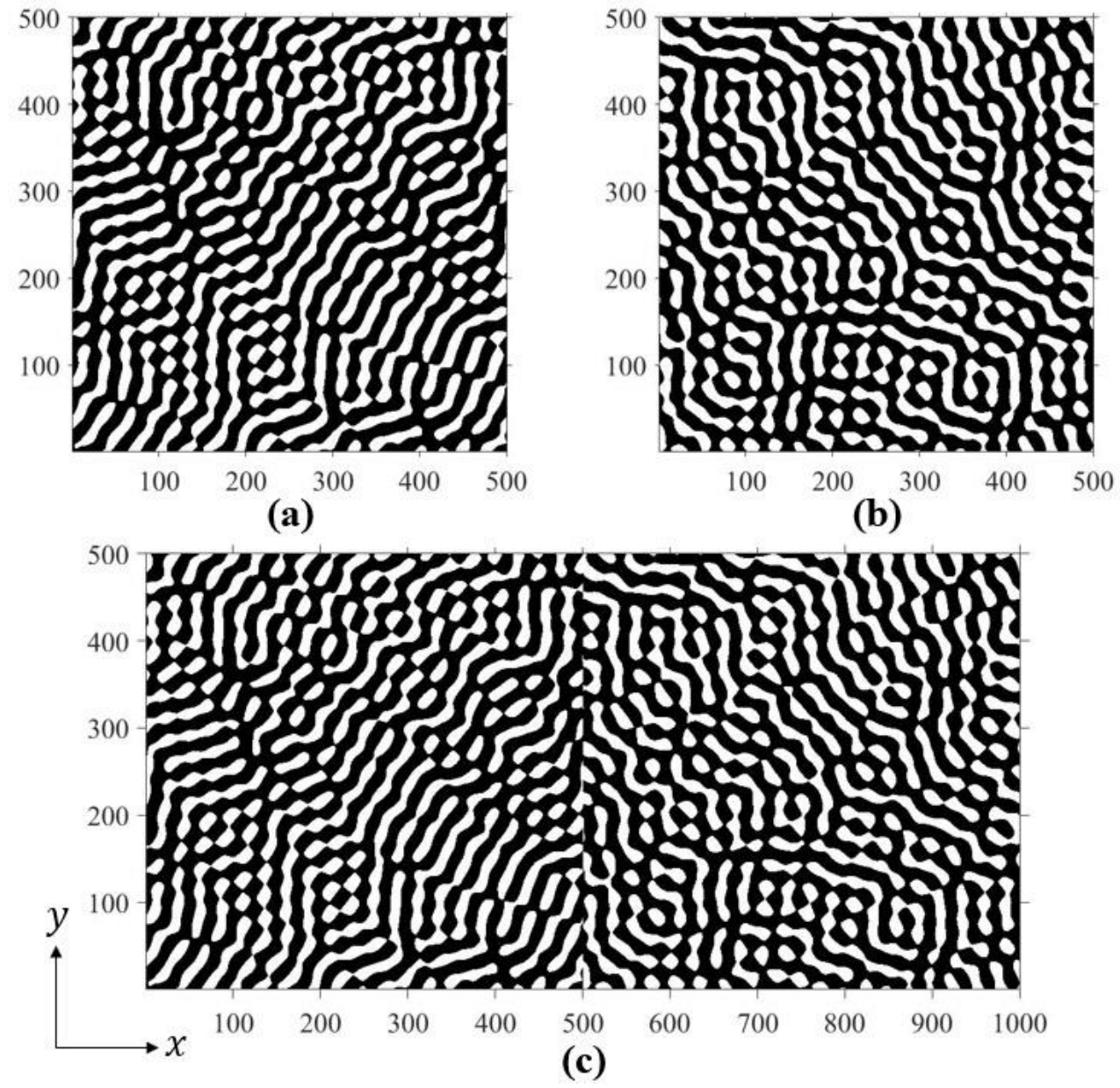

**Figure 5 Illustration of ill-connected interfaces:** When connecting two spinodoid microstructures in **(a)** and in **(b),** both of which have a material volume fraction of 50% and are discretized by $500 \times 500$ pixels, their interface at $x = 500$ shows ill-connected interface with discontinuous and isolated materials as seen in **(c)**.

To enable a smooth connection at the interfaces between disparate microstructures, we propose a two-step approach. Firstly, we apply an interface interpolation function to the phase fields of two neighboring microstructures. Our interpolation function is:

$$\varphi_{12}^{\text{int}}(\boldsymbol{X}) = \left(1-\lambda(x)\right)\varphi_1(\boldsymbol{X}) + \lambda(x)\varphi_2(\boldsymbol{X}), \text{ and} \tag{8}$$

$$\lambda(x) = \frac{e^{-\zeta(x-2l)^2}}{e^{-\zeta(x-l)^2} + e^{-\zeta(x-2l)^2}}, \tag{9}$$

where $\varphi_{12}^{\text{int}}$ is the interpolated phase fields, while $\varphi_1$ and $\varphi_2$ are the phase fields of to-be-connected microstructures. $\lambda(x)$ is the interpolation function depending on the axis value $x$ (horizontal or vertical axes). While $l$ represents the size of the microstructure, e.g., $l$=500 in Figure 5, $\zeta$ is a user-defined constant that decides the size of interpolation regions. For example, the interpolation function for the two microstructures of Figure 5(c) is given in Figure 6 with $\zeta$=5e-5.

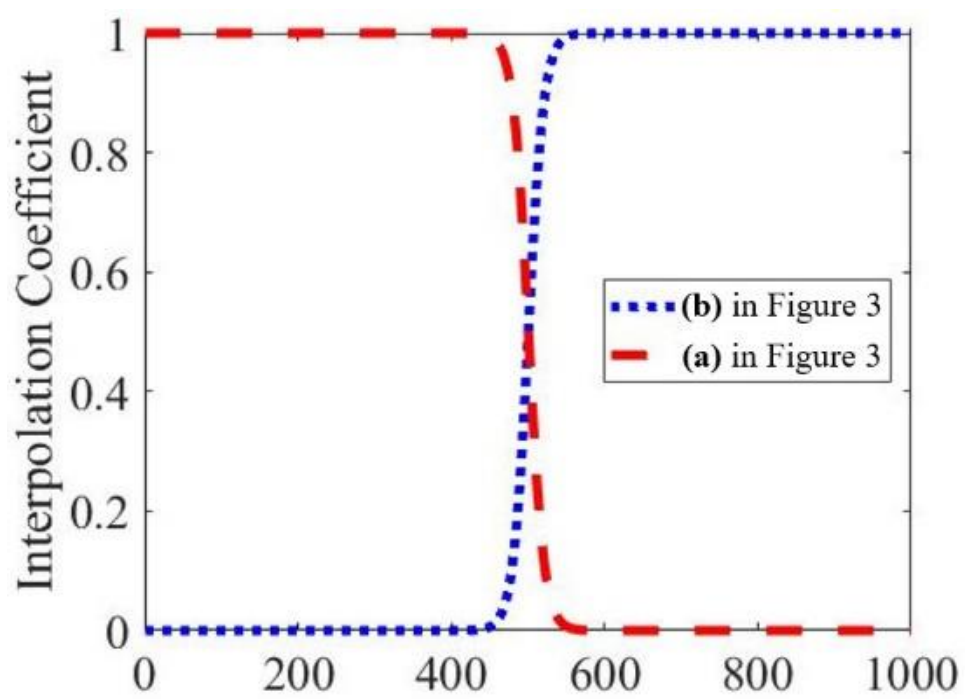

**Figure 6 Interface interpolation function:** Two interpolation functions are applied to the phase fields of the two microstructures in Figure 5, respectively, where the user-defined constant is *ζ=5e-5* in Equation (9).

We note that the smaller the value of $\zeta$, the larger interpolation regions are used within the two microstructures. For example, when we set $\zeta$ as 5e-4, 5e-5, and 5e-6, the interpolation results are shown in Figure 7(a)-(c), respectively.

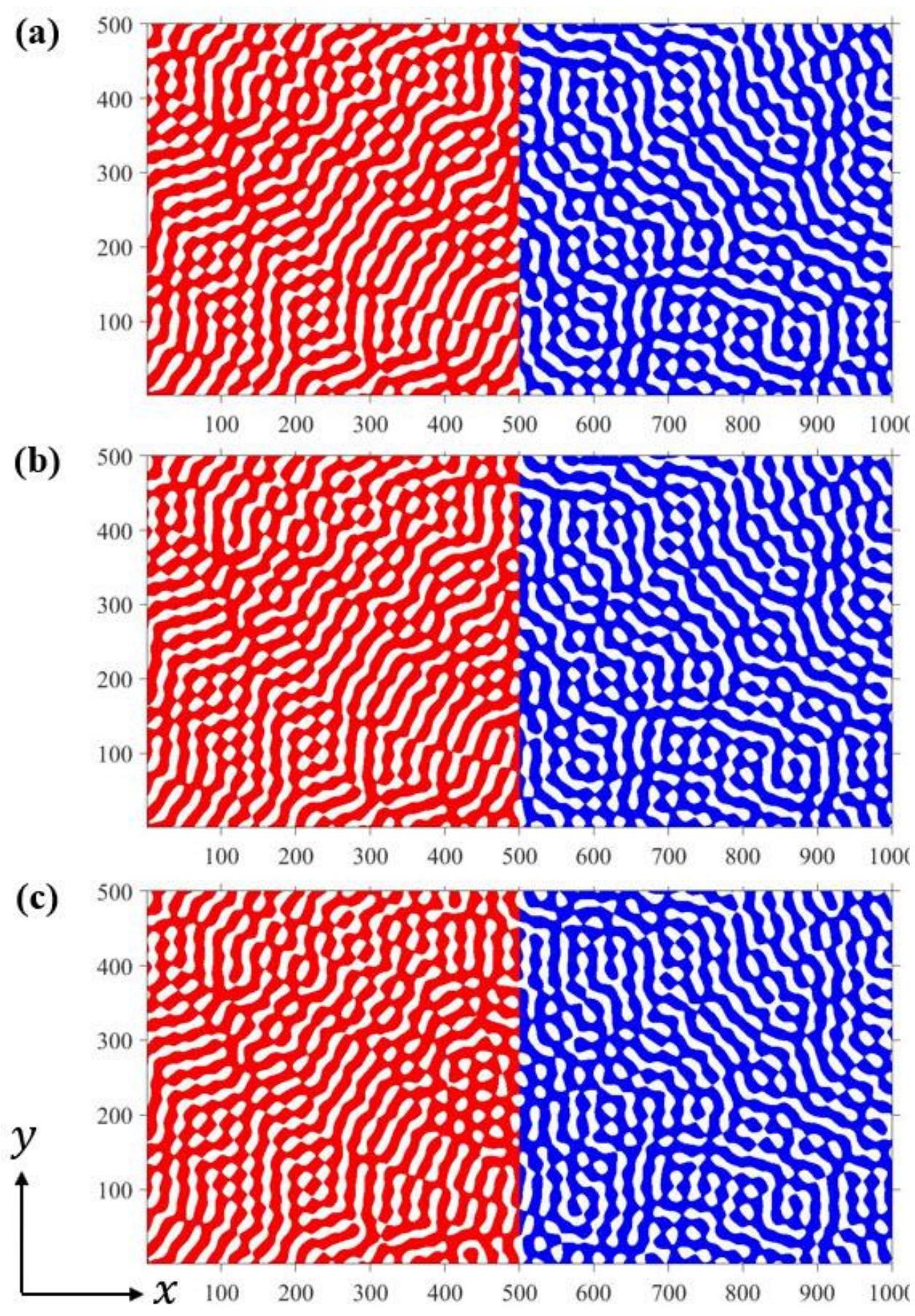

**Figure 7 Demonstration of the effects of interpolation functions:** As decreasing the value of $\zeta$ of Equation (9) from 5e-4 in **(a)**, to 5e-5 in **(b)**, and 5e-6 in **(c)**, we observe smoother transitions between the two microstructures at the interface at $x = 500$ (which represents pixel index rather than a length unit). Here, we use color codes to indicate two different spinodoid microstructures, i.e., they do not represent the types of spinodoids.

It is evident that compared with the non-interpolation microstructure of Figure 5(c), the interface becomes more and more well-connected as we reduce the value of $\zeta$ when applying the interpolation function. However, when $\zeta$ becomes too small, the interpolation region would significantly change the microstructure's interior region far away from the interface, as in Figure 7(c). For this reason, we choose $\zeta$=5e-5 as the default value for interpolation.

We note that the interface interpolation function works well when the neighboring microstructures have similar material volume fractions. To enhance the interface connection between microstructures of significantly different volume fractions, we introduce an interface amplifying function in our second step as follows:

$$\varphi_{12}^{\text{amp}}(\boldsymbol{X}) = \eta\left[1+\gamma(x)\right]\varphi_{12}^{\text{int}}(\boldsymbol{X}), \text{ and} \tag{10}$$

$$\gamma(x) = \begin{cases} \lambda(x), & \text{if } x \le l \\ 1-\lambda(x), & \text{otherwise} \end{cases} \tag{11}$$

where $\varphi_{12}^{\text{amp}}$ is the amplified phase fields from the interpolation $\varphi_{12}^{\text{int}}$ in Equation (8), and $\gamma(x)$ is the amplifying function depending on the value $x$ along the interpolation axis. We note that $\eta$ is the amplifying coefficient that controls the magnifying ratio for the phase fields on the boundary interface. For instance, we demonstrate the interface amplifying function as in Figure 8 with $\eta = 1$.

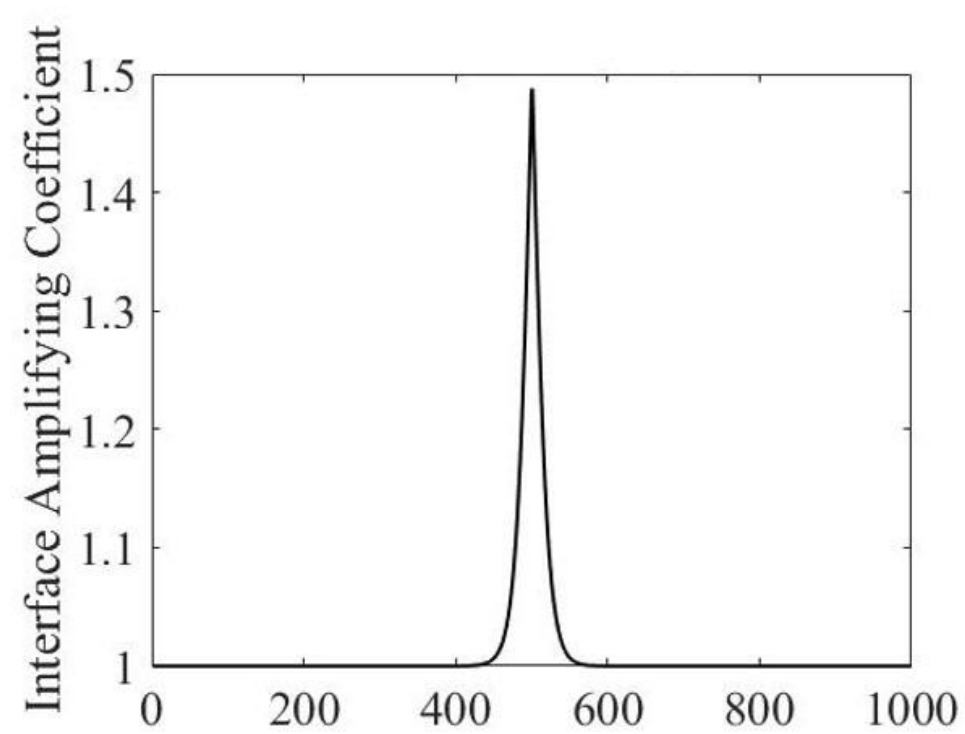

**Figure 8 Interface amplifying function:** Interface amplifying coefficients are applied along the interpolation direction. In proximity to the interface at $x = 500$, the phase fields of the two neighboring microstructures are multiplied by a magnifying coefficient.

To demonstrate the effects of the amplifying function, we set $\eta$ as three values 0 (no amplifying effect), 1, and 10, and show their interpolated interfaces for two neighboring microstructures with 50% and 80% volume fractions in Figure 9(a)-(c), respectively. We observe that as we increase $\eta$, the interfaces becomes more and more well-connected. We therefore choose $\eta = 10$ as the default value in this work.

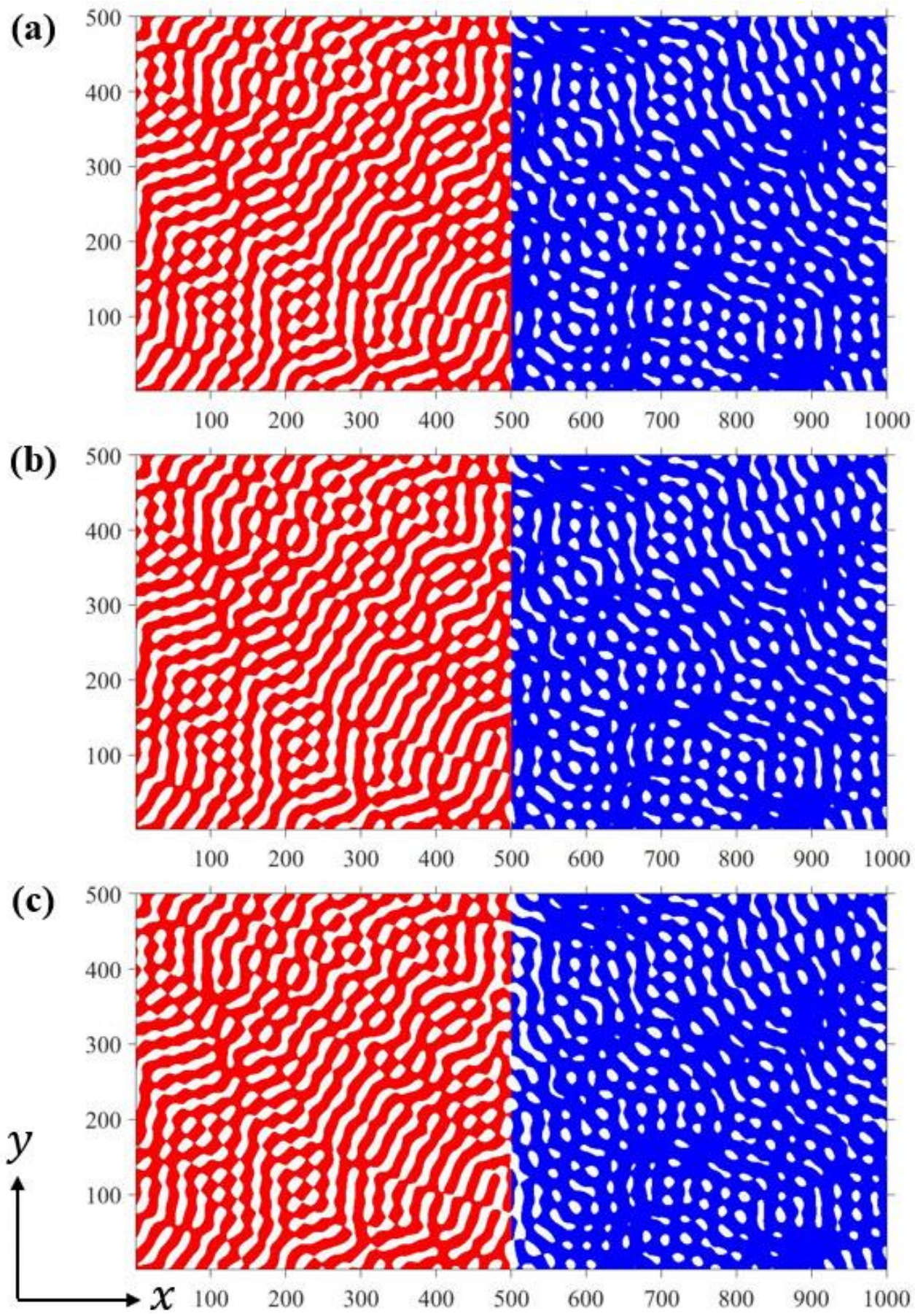

**Figure 9 Demonstration of the effect of the amplifying function:** As increasing the value of the amplifying parameter $\eta$ of Equation (10) from 0 in **(a)**, to 1 in **(b)** and 10 in **(c)**, we observe well-connected boundary interfaces at $x = 500$ for the two microstructures with significantly different material volume fractions, i.e., 50% and 80%.

### 3.3. Effective Elastic Properties of Spinodoid Microstructures

After reconstructing the spinodoid microstructures, we aim to compute their effective elastic tensors. Here, we follow the computational homogenization scheme in [27] to use Finite Element Analysis (FEA) to approximate the homogenized linear elastic stiffness tensor $C^H_{ijkl}$ under periodic boundary conditions:

$$C^H_{ijkl} = \frac{1}{|\boldsymbol{X}|}\sum_{e=1}^{N}\left(\boldsymbol{u}_e^{A(ij)}\right)^T \boldsymbol{k}_e \boldsymbol{u}_e^{A(kl)} \;, \tag{12}$$

where $\boldsymbol{u}_e^{A(kl)}$ and $\boldsymbol{u}_e^{A(ij)}$ are the elemental displacement solutions corresponding to unit test train fields and admissible displacement fields, respectively, and $\boldsymbol{k}_e$ is the element stiffness matrix. For our 2D case, Equation (12) can be simplified by noting $11 \rightarrow 1$, $22 \rightarrow 2$, and $12 \rightarrow 3$, as an expanded form as:

$$C^H_{ij} = \frac{1}{|\boldsymbol{X}|}\sum_{e=1}^{N} q_e^{(ij)} \text{ , and} \tag{13}$$

$$q_e^{(ij)} = \left(\boldsymbol{u}_e^{A(i)}\right)^T \boldsymbol{k}_e \left(\boldsymbol{u}_e^{A(j)}\right), \tag{14}$$

where $q_e^{(ij)}$ are the element mutual energies from the test and admissible element displacement solutions, $\boldsymbol{u}_e^{A(i)}$ and $\boldsymbol{u}_e^{A(i)}$, respectively.

For the three types of spinodoids considered in this work, isotropic spinodoids have infinitely many planes of symmetry, while monoclinic and orthotropic spinodoids have one and two planes of symmetry, respectively. Therefore, we can further reduce the number of elastic constants of their homogenized stiffness tensor matrix to two, six, and four, respectively.

We note that the homogenization approach of the Equation (12) is developed for periodic boundary conditions, but not for the aperiodic boundary conditions after the interface smoothing in Section 3.2. In this work, we hypothesize that the boundary smoothing procedure of Section 3.2 may not significantly change the microstructural effective elastic properties. To test our hypothesis, we use the interpolation and amplifying functions to smoothen the interfaces of the four microstructures in Figure 10, and then compute their homogenized elastic constants.

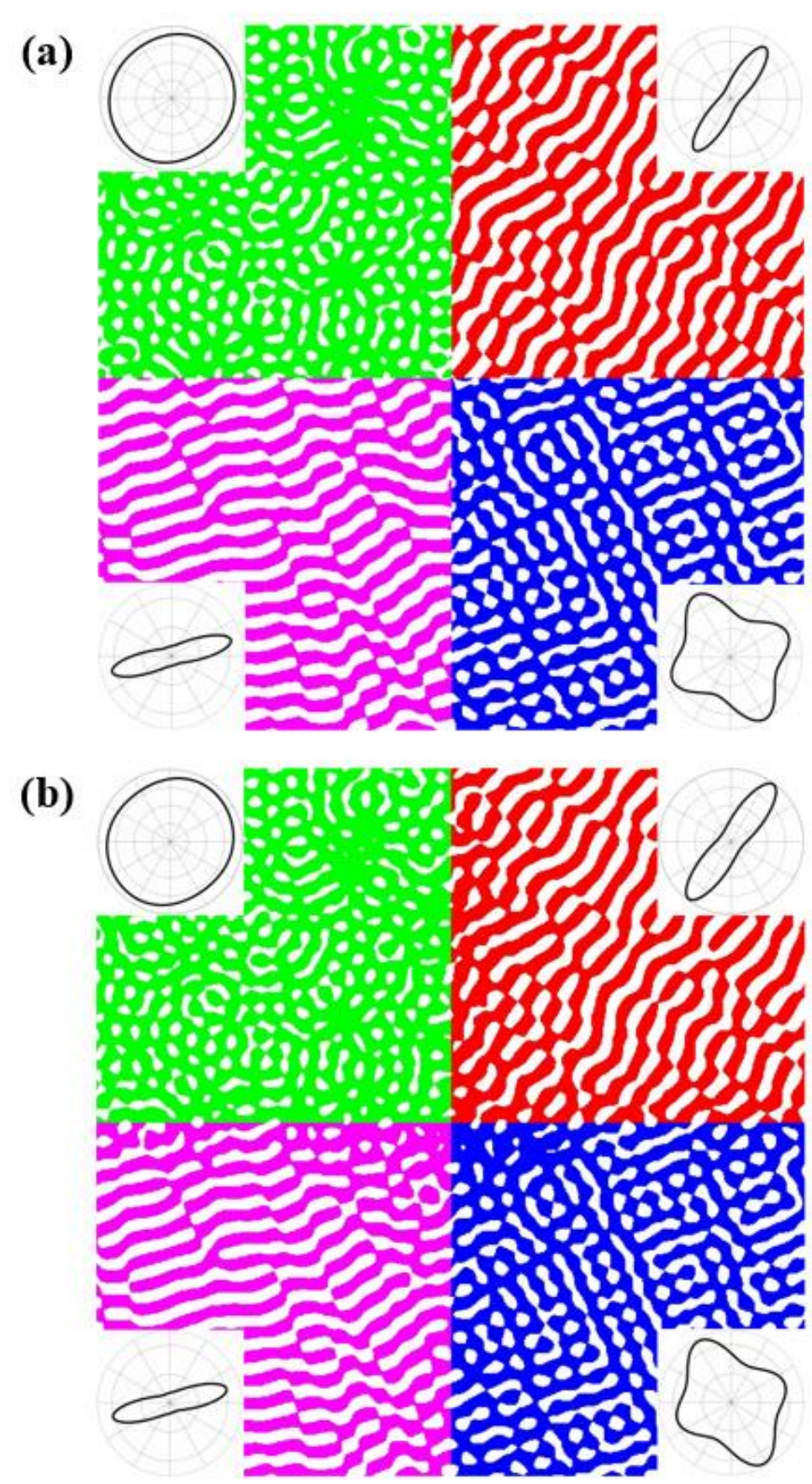

**Figure 10 Impacts of interface connection on effective elastic properties:** Boundary interface connections are compared before and after applying the interface interpolation and amplifying functions in both horizontal and vertical directions on four different spinodoid microstructures, where homogenized elastic surfaces show directional dependency.

As in Figure 10, there are one isotropic (color coded as green), one orthotropic (blue), and two monoclinic spinodoids with different directional anisotropy (red and purple). By comparing Figure 10(a) and (b), it is obvious that their interfaces become better connected after interface interpolation. We use the Relative Root Mean Squared Error (RRMSE) to quantify the difference before and after enforcing interface connection as:

$$RRMSE = \sqrt{\frac{\frac{1}{n}\sum_{i=1}^{n}\left(C_i - \tilde{C}_i\right)^2}{\sum_{i=1}^{n}\left(\tilde{C}_i\right)^2}}, \quad (15)$$

where $C_i$ and $\tilde{C}_i$ are the Voigt notations of the homogenized elastic tensor constants before and after enforcing the microstructural interface connection. The RRMSEs for the four microstructures are shown in Table 1.

We note that even though the boundary smoothing procedure does not seem to result in significant change in the microstructural effective properties as shown in Table 1 in this example, we emphasize that the influence of the smoothing procedure on the topologically optimized structures depends on many factors, such as the morphologies of local and neighboring microstructures, values of hyper-parameters in interface interpolation and amplifying functions, and the type of boundary conditions used in homogenization. We will need to conduct a systematic study to investigate their impact on the effective elastic properties of the spinodal architected materials in the future.

**Table 1 RRMSE of elastic constants:** RRMSEs compare the homogenized elastic constants before and after enforcing boundary interface connections in Figure 10.

| Spinodoids | Green | Red | Purple | Bule |
|---|---|---|---|---|
| RRMSE | 0.0217 | 0.0148 | 0.0370 | 0.0092 |

### 3.4. Data Repository and Gaussian Process

After the microstructure reconstruction discussion from the previous section, we proceed to use the Design of Experiment (DoE) [41] in this section to generate datasets of reconstructed spinodoid samples as demonstrated in Figure 2. The datasets are used for training Gaussian Process (GP) that predicts spinodoid effective elastic moduli based on their microstructural descriptors.

We would like to emphasize that the three types of spinodal microstructures have different numbers of design variables. Specifically, the isotropic spinodoids have two design variables (therefore, the dimension of inputs for GP is two), the monoclinic ones have four variables (input dimension is four), and the orthotropic ones have five variables (input dimension is five). A rule-of-the-thumb is that the number of samples should be no less than ten times the input dimension, that is, 20, 40, and 50 samples for the isotropic, monoclinic and orthotropic spinodoids, respectively. In our work, we generate three datasets in total. In other words, we generate one dataset for each of the three types of spinodoid architected materials (see Section 3.1). Each dataset contains 600 (500 for training and 100 for testing) spinodoid microstructure samples. The values of each dataset are the values of spinodoid geometric descriptors.

For isotropic spinodoids, the two geometric descriptors are the microstructure density $\rho_m$ and the frequency $k$ from the target phase field (signals in LTI) as in Figure 3(a), and their ranges are shown below:

$$\begin{cases} 0 \le \rho^m \le 1 \\ 10 \le k \le 30. \end{cases} \quad (16)$$

For the monoclinic spinodoids, the ranges of the two additional descriptors, i.e., the anisotropic index $\alpha^{\text{mon}}$ and the frequency rotational angle $\gamma$, are:

$$\begin{cases} 0 \le \alpha^{\text{mon}} \le 1 \\ 0 \le \gamma \le \pi. \end{cases} \quad (17)$$

For orthotropic spinodoids, the geometric descriptors include not only the microstructure material density $\rho^m$ and the frequency $k$ from the target phase field from Equation (16), but also the frequency rotational angle $\gamma$, and two anisotropic indices $\alpha_1^{\text{ort}}$ and $\alpha_2^{\text{ort}}$. Their ranges are as follows:

$$\begin{cases} 0 \le \gamma \le \pi \\ \sqrt{2}/2 \le \alpha_1^{\text{ort}} \le 1 \\ \sqrt{2}/2 \le \alpha_2^{\text{ort}} \le 1. \end{cases} \quad (18)$$

We would like to note that limiting the ranges of design variables can reduce our design space, therein preventing from finding better local optimum or global optimum if possible. However, our reduction of the range of design variables is to preserve spinodal morphologies generated from our SDF-based reconstructions. For instance, in our experiments, we found that, as the anisotropic index increases towards the range of [0.9, 1.0], the reconstructed morphology starts to resemble that of truss or beam-like metamaterials. Additionally, we chose microscale density in the range of [0.3, 0.7] to improve bi-continuity of the two material phases in spinodoids. This is because there is no theoretical proof (to the best knowledge of the authors) that SDF-reconstructed 2D spinodoids can always maintain bi-continuity at all microscale volume fractions in the range of [0, 1.0]. Particularly, if we increase the microscale volume fraction above 0.7 or reduce it to below 0.3, we are more likely to observe material phase discontinuity than that in the range of volume fraction of [0.3, 0.7].

We note that we used Sobol sequence to draw samples from the input space to create random SDF parameters followed by spinodal microstructure reconstructions. We demonstrate the scattering plots of the SDF parameters in the input space as in Figure 11. Compared to classic DoE approaches like Latin Hypercube, Sobol sequence improves the sampling diversity within input space and allows for sequential sampling (not overlapping with previous samples).

Once the datasets are built, we use the SDF-based microstructure reconstruction to rebuild spinodoid microstructures following Section 3.1, and adopt the homogenization method in Section 3.3 to calculate the effective elastic constants for all microstructure samples, i.e., two Lame constants $\mu$ and $\lambda$ for isotropic spinodoids, six moduli components $[C_1, C_2, C_3, C_4, C_5, C_6]$ for monoclinic spinodoids, and four moduli constants $[C_1, C_2, C_3, C_4]$ for orthotropic spinodoids. We demonstrate the sample distributions of the effective elastic constants for each of the three datasets, as shown in Figure 12. We should note that the scattering plots in Figure 12 demonstrate the distributions of output variables of our GP (elastic tangent components), and they do not represent sampling density in the input (parameter) space which are shown in Figure 11. Furthermore, we normalize the values of all spinodoid descriptors to the range of $[0, 1]$ to be compatible with the NNs-based parameterization scheme as discussed in Section 3.5.

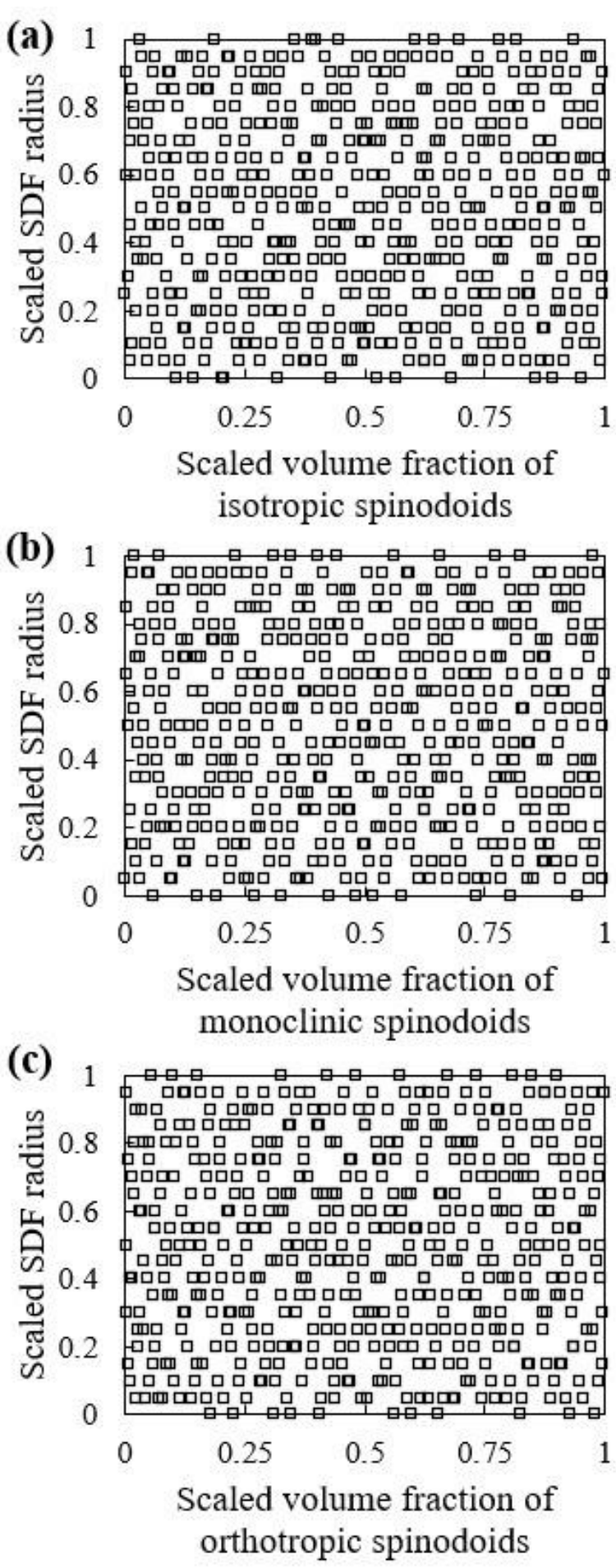

**Figure 11 Scattering plots of spinodoid design parameters in the input space of GP:** The scattering plots between microstructural volume fractions and SDF radii are illustrated for the 500 training samples for **(a)** isotropic, **(b)** monoclinic, and **(c)** orthotropic architected materials.

In our work, we only generate three types of spinodoids. Our spinodoids were defined based on the number of symmetric planes in the architected materials, that is, isotropic spinodoids with infinite symmetric planes, monoclinic spinodoids with one symmetric plane, and orthotropic spinodoids with two orthogonal planes. These three spinodoid types cannot cover the entire spectrum of spinodoids. We could extend our spinodoid material system to include more symmetric planes (like 4 or 8 symmetric planes) or un-orthogonal symmetric planes in our future study. However, we note that choosing the number of symmetric planes will depend on target material systems for reconstruction.

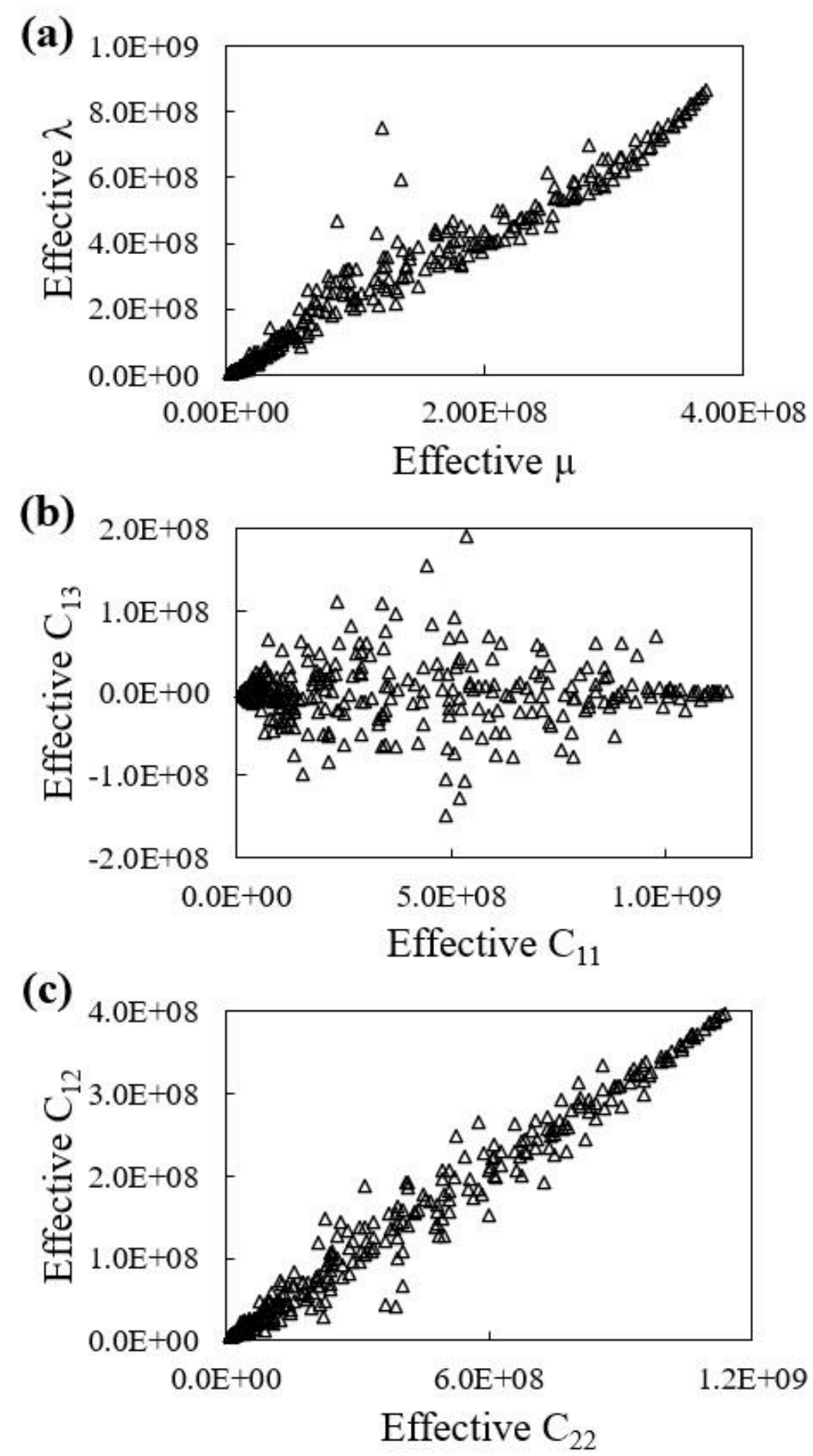

**Figure 12 Distributions of effective elastic properties of training spinodoid samples:** The effective elastic properties are computed for 500 training data points for each of the three spinodoid types, i.e., **(a)** isotropic, **(b)** monoclinic, and **(c)** orthotropic architected materials.

For GP training, we trained three multi-response GPs. Specifically, we generated one multi-response GP for each of the three spinodoid types, and our GP outputs include multiple elastic modulus components.

In each dataset, we randomly split the 600 samples into 500 samples for training and 100 samples for testing. We proceed to train a GP for each of the three datasets to emulate the relation between spinodoid descriptors and their effective elastic constants.

GP, widely used for emulation [42], [43], [44], [45], approximates each scalar quantify of interest $q$ as a realization of the random field $Q$ over the $d_s$ dimensional input space with the input variables $\boldsymbol{s} = [s_1, s_1, \dots, s_{d_s}]^T$. Specifically, the random field $Q$ can be written as:

$$Q(\boldsymbol{s}) = \boldsymbol{m}(\boldsymbol{s})^T \boldsymbol{\beta} + \varepsilon(\boldsymbol{s}), \tag{19}$$

where $\boldsymbol{m}^T(\boldsymbol{s})\boldsymbol{\beta}$ is the mean as in Equation (20) and $\varepsilon(\boldsymbol{s})$ is a zero-mean GP with the covariance function defined as in Equation (21):

$$\mathrm{E}\left[Q(\boldsymbol{s})\right] = \boldsymbol{m}(\boldsymbol{s})^T \boldsymbol{\beta}, \text{ and} \tag{20}$$

$$Cov\left[Q(\boldsymbol{s}),\ Q(\boldsymbol{s}')\right] = Cov\left[\varepsilon(\boldsymbol{s}),\ \varepsilon(\boldsymbol{s}')\right] = c\left(\boldsymbol{s},\ \boldsymbol{s}'\right), \tag{21}$$

where $\boldsymbol{m}(\boldsymbol{s}) = [m_1(\boldsymbol{s}), m_2(\boldsymbol{s}), \dots, m_u(\boldsymbol{s})]^T$ is a vector with $u$ known basis functions, and $\boldsymbol{\beta} = [\beta_1, \beta_2, \dots, \beta_u]^T$ is a vector of $u$ unknown parameters. For simplicity, by assuming $u = 1$ and $\boldsymbol{m}(\boldsymbol{s}) = [1]^T$, our prior mean is a constant of the value of $\beta$.

We note that in Equation (21), a kernel trick is applied to replace the covariance function $Cov[Q(\boldsymbol{s}), Q(\boldsymbol{s}')]$ of the random field $Q$ as the covariance of the input variables $\boldsymbol{s}$. When we choose the kernel $c(\boldsymbol{s}, \boldsymbol{s}')$ to follow the Gaussian covariance, it can be expressed as a correlation function $r(\boldsymbol{s}, \boldsymbol{s}')$ as:

$$c\left(\boldsymbol{s},\ \boldsymbol{s}'\right) = \sigma^2 r\left(\boldsymbol{s},\ \boldsymbol{s}'\right), \text{ and} \tag{22}$$

$$\begin{aligned} r\left(\boldsymbol{s},\ \boldsymbol{s}'\right) &= exp\left[-\sum_{i=1}^{d_s} 10^{w_i} \left(s_i - s_i'\right)^2\right] \\ &= exp\left[\left(\boldsymbol{s} - \boldsymbol{s}'\right)^T \boldsymbol{\Omega}_s \left(\boldsymbol{s} - \boldsymbol{s}'\right)\right], \end{aligned} \tag{23}$$

where $\sigma^2$ is the prior variance of the random field $Q(\boldsymbol{s})$. Additionally, $\boldsymbol{w} = [w_1, w_2, \dots, w_{d_s}]^T$ is the vector of correlation parameters, also called roughness parameters, controlling the smoothness of the random field $Q(\boldsymbol{s})$, and $\Omega_{\mathbf{s}} = diag(10^{\boldsymbol{w}})$. The correlation function $r(\boldsymbol{s}, \boldsymbol{s}')$ can be viewed as a weighted distance between the two inputs $\boldsymbol{s}$ and $\boldsymbol{s}'$.

To estimate the model parameters $\boldsymbol{\beta}$, $\sigma^2$ and $\boldsymbol{w}$, we follow the practice of Maximum Likelihood Estimation (MLE) [42], which is equivalent to maximizing the logarithm of the likelihood as:

$$\left[\hat{\boldsymbol{\beta}},\ \hat{\sigma}^2,\ \hat{w}\right] = \underset{\boldsymbol{\beta}, \sigma^2, \boldsymbol{w}}{\arg\max} \left\{\log\left[L(\boldsymbol{\beta}, \sigma^2, \boldsymbol{w} | \boldsymbol{q})\right]\right\}, \tag{24}$$

where $\hat{\bullet}$ indicates the estimated parameters, and the likelihood $L = p(\boldsymbol{q}|\boldsymbol{\beta}, \sigma^2, \boldsymbol{w})$ is the probability of the observations, i.e., the quantity of interests $\boldsymbol{q} = [q(\boldsymbol{s}^{(1)}, \boldsymbol{s}^{(2)}, \dots, \boldsymbol{s}^{(i)} \dots, \boldsymbol{s}^{(r)})]^T$ at $r = 500$ sampling points where $\boldsymbol{s}^{(i)}$ indicate the $i^{th}$ sample, conditioning on GP model parameters. Following profiling [46], the predicted quantity of interests $\hat{q}(\boldsymbol{s}^{test})$ at an unsampled testing point $\boldsymbol{s}^{test}$ can be computed as:

$$\hat{q}(\boldsymbol{s}^{test}) = \boldsymbol{m}^T(\boldsymbol{s}^{test})\hat{\boldsymbol{\beta}} + \boldsymbol{c}^T(\boldsymbol{s}^{test})\boldsymbol{C}^{-1}\left(\boldsymbol{q} - \boldsymbol{M}\hat{\boldsymbol{\beta}}\right), \tag{25}$$

where $\boldsymbol{c}(\boldsymbol{s}^{test})$ is an $r \times 1$ vector with the $i^{th}$ element as $c(\boldsymbol{s}^{(i)}, \boldsymbol{s}^{test})$, $\boldsymbol{C}$ is a $r \times r$ matrix with the element at $(i, j)$ as $c(\boldsymbol{s}^{(i)}, \boldsymbol{s}^{(j)})$, and $\boldsymbol{M}$ is a $r \times u$ matrix with the $i^{th}$ row as $\boldsymbol{m}^T(\boldsymbol{s}^{(i)})$.

We test prediction accuracy on the 100 testing samples for each of the three datasets (each dataset corresponds to one spinodoid type) as shown in Figure 13. Specifically, we demonstrated the GP's prediction accuracy in the scattering plot by comparing ground truth with predicted elastic moduli. We computed the RRMSE of elastic moduli components and showed that they were in the range of [2.59e-3, 8.05e-3]. RRMSE is a normalized unit-less error metric which indicates the relative error with respect to the scale of actual data. Since our trained GPs have RRMSE smaller than 1%, we believe that all three GPs show high prediction accuracy, and they are ready for on-line deployment in the multiscale TO that we discuss in Section 3.5.

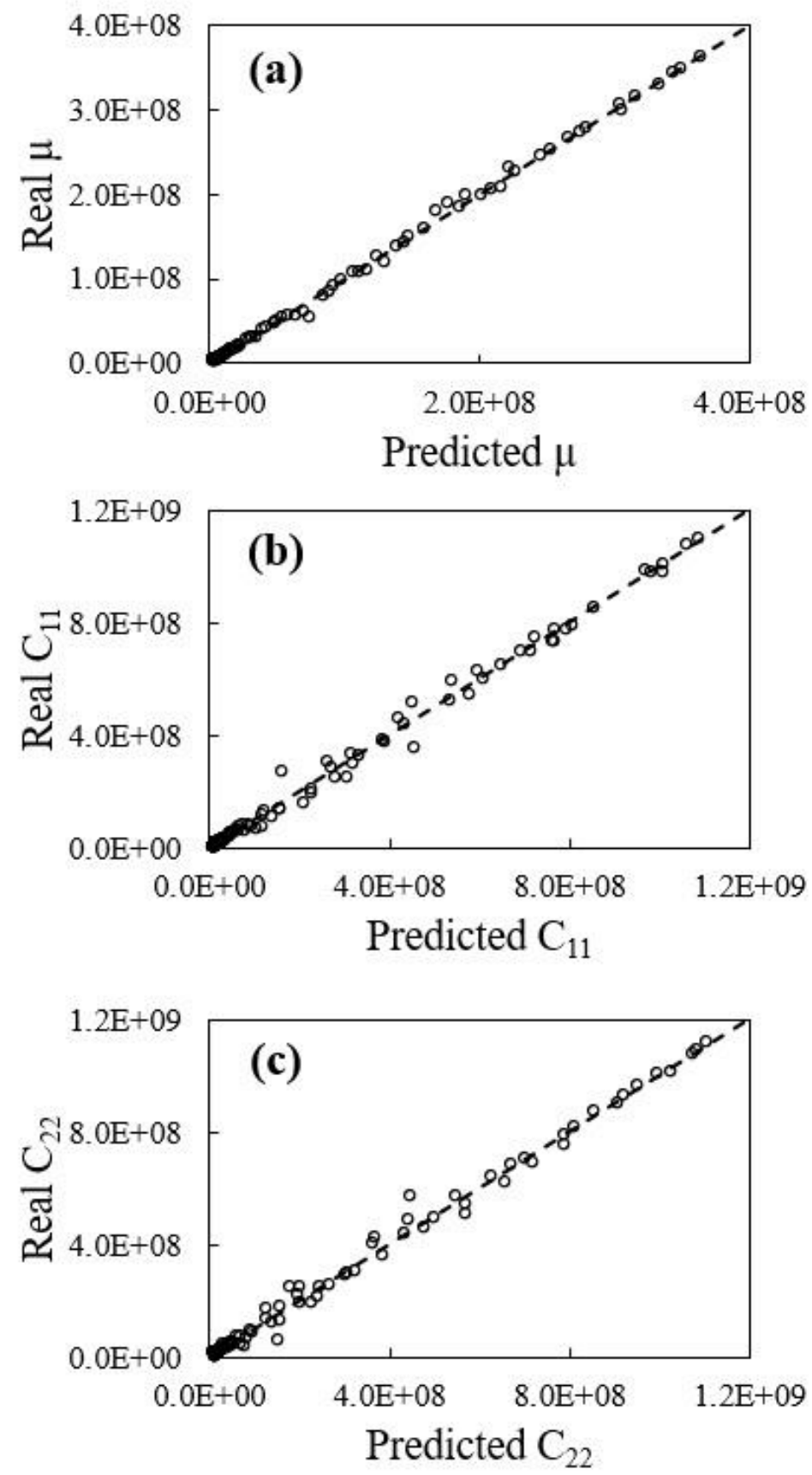

**Figure 13 GPs' prediction accuracy on the effective elastic constants:** GPs are used to predict elastic constants for the 100 testing data points for each of the three datasets, for the **(a)** isotropic, **(b)** monoclinic, and **(c)** orthotropic spinodoids, with the prediction RRMSEs of 8.05e-3, 2.59e-3, and 3.37e-3, respectively.

### 3.5. Neural Networks-Based Topology Optimization

In this section, we use the TO approach to solve the multiscale spinodoid design. Following the bottom-up TO approach [47], our multiscale TO is essentially reduced to a structure-level single-scale TO. For this single scale TO, we use $\boldsymbol{\Theta}$ to include all spinodoid design parameters as:

$$\boldsymbol{\Theta} = \left\{ \boldsymbol{T},\ \boldsymbol{\rho}^{M},\ \boldsymbol{\rho}^{m},\ \boldsymbol{k},\ \boldsymbol{\gamma},\ \boldsymbol{\alpha}^{mon},\ \boldsymbol{\alpha}_1^{ort},\ \boldsymbol{\alpha}_2^{ort} \right\}, \tag{26}$$

The spinodoid TO can be formulated as:

$$\text{Minimize} \quad J(\boldsymbol{\Theta}) = \boldsymbol{F}^T \boldsymbol{U} \tag{27}$$

$$\text{S.T.} \quad \begin{cases} \boldsymbol{K}(\boldsymbol{\Theta})\mathbf{U} = \mathbf{F} \\ g = \left( \sum_{e=1}^{d_e} \rho_e \mathrm{A}_e \right) \Big/ \left( \rho^t \sum_{e=1}^{d_e} \mathrm{A}_e \right) - 1 \le 0 \\ \rho_e = \rho_e^M \rho_e^m \\ 0 \le \rho_e^M \le 1 \\ 0 \le \rho_e^m \le 1 \\ t_e^{iso} + t_e^{mon} + t_e^{ort} = 1, \end{cases} \tag{28}$$

where $\boldsymbol{T}$ is a $d_e \times 3$ matrix indicating the composition of spinodoid types per macroscale element. For example, $\boldsymbol{t}_e = [10\%, 60\%, 30\%]$ is the $e^{th}$ row of the $\boldsymbol{T}$, indicating the $e^{th}$ macroscale element contains 10%, 60%, and 30% of isotropic, monoclinic, and orthotropic spinodoids, respectively. In other words, we convert the categorical variables of spinodoid types to continuous variables, and enforce their summation as one. That is, we do not restrict the design to a specific type of spinodoid; rather, all three types are permitted to coexist for each macroscale element.

In addition, $\rho_e^M$ and $\rho_e^m$ represent the macroscale and microstructural densities, respectively, and their multiplication results in the composed (actual) density $\rho_e$ for the $e^{th}$ element. For material interpolation, we penalize $\rho_e^M$ and $\boldsymbol{t}_e$ by following the Solid Isotropic Material with Penalization (SIMP) scheme as:

$$\tilde{\boldsymbol{C}}_e = \sum_{ty=1}^{3} \left[ \left( \boldsymbol{C}_e^{ty}(\boldsymbol{\Theta}) - \boldsymbol{C}_{\min}^{ty}(\boldsymbol{\Theta}) \right) \left( \rho_e^M t_e^{ty} \right)^p + \boldsymbol{C}_{\min}^{ty}(\boldsymbol{\Theta}) \right], \tag{29}$$

where the penalized elastic moduli $\tilde{\boldsymbol{C}}_e$ per element is the summation of the penalized moduli of each spinodoid type. $\boldsymbol{C}_e^{ty}$ represents the predicted effective elastic moduli from GP prediction for the three types of spinodoids, as discussed in Section 3.4. $ty = 1,2,3$ corresponds to the isotropic, monoclinic, and orthotropic spinodoids, respectively. $\boldsymbol{C}_{min}^{ty} = 1\mathrm{e}^{-6}\boldsymbol{C}_e^{ty}$ is set as the lower bound of the elastic moduli to avoid numerical singularity. We follow the continuation approach [48] to penalize both the $\rho_e^M$ and $\boldsymbol{t}_e$, such that $\rho_e^M$ is penalized to either 0 or 1, and one of the three spinodoid types in $t_e^{ty}$ is penalized to 1, as optimization proceeds towards convergence. The penalized elastic moduli $\tilde{\boldsymbol{C}}_e$ can form the element stiffness matrix for the linear elastic analysis as:

$$\boldsymbol{K}_e = \int \boldsymbol{B}^T \tilde{\boldsymbol{C}}_e \boldsymbol{B} d\Omega_e\ , \tag{30}$$

We adopt the NNs-based TO model introduced in [49] and extend it for our spinodoid design. We assume that the dimensions of the input, output, and the hidden layers as two, ten and 20, respectively, as in Figure 14. We note that this shallow NNs work as an implicit function mapping the coordinates of macro-elements to the design parameters of associated spinodoid microstructures, including spinodoid type, density, frequency, anisotropic index, and rotational angle, as discussed in Sections 3.1 and 3.2.

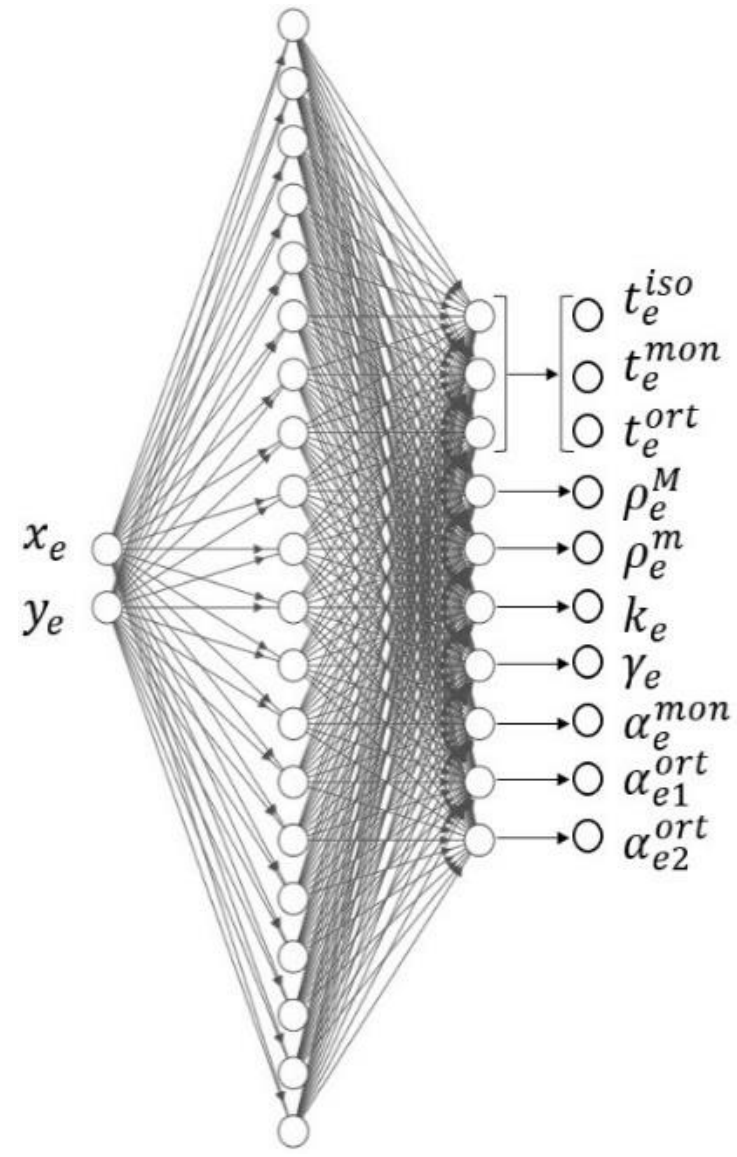

**Figure 14 NNs-based design re-parameterization:** The shallow NNs contain one hidden layer of 20 neurons, two input neurons and ten output neurons. While the input layer reads in the 2D coodinates of macroscale element centroids, the output layer outputs spinodoid design parameters.

By using NNs, we implicitly represent the spinodoid design parameters $\boldsymbol{\Theta}$ as the functions of the NNs' weight parameters $\boldsymbol{\tau}$ as:

$$\boldsymbol{\Theta} = \boldsymbol{\Theta}(\tau), \tag{31}$$

We can therefore parameterize the TO formulation from Equations (26)-(28) as follows:

$$\text{Minimize} \quad J(\boldsymbol{\tau}) = \boldsymbol{F}^T \boldsymbol{U} \tag{32}$$

$$\text{S.T.} \quad \begin{cases} \boldsymbol{K}(\boldsymbol{\tau})\mathbf{U} = \mathbf{F} \\ g = \left( \sum_{e=1}^{d_e} \rho_e(\boldsymbol{\tau}) \mathrm{A}_e \right) \Big/ \left( \rho^t \sum_{e=1}^{d_e} \mathrm{A}_e \right) - 1 \le 0, \end{cases} \tag{33}$$

where all spinodoid design parameters are functions of NNs' weight parameters as shown in Figure 14.

We apply a SoftMax function on the three output neurons of Figure 14 corresponding to the spinodoid types, such that the condition $t_e^{iso} + t_e^{mon} + t_e^{ort} = 1$ is automatically satisfied. We apply sigmoid functions to map the other outputs to the range of $[0, 1]$, such that the constraints of the macro- and micro-scale material densities are both satisfied. Additionally, for the sake of material connectivity and differentiation between three spinodoid types, we linearly map the range of the microscale density and anisotropic indices to the following ranges:

$$\begin{cases} 0.3 \le \rho_e^m \le 0.7 \\ 0.5 \le \alpha_e^{mon} \le 0.9 \\ 0.5 \le \alpha_{e1}^{ort} \le 0.9 \\ 0.5 \le \alpha_{e2}^{ort} \le 0.9, \end{cases} \tag{34}$$

where $\rho_e^m$ is assumed to vary in the range of $[0.3, 0.7]$. It is not subject to penalization, so that the microscale density could vary without penalization in the scaled range. In addition, the anisotropic indices are assumed smaller than 0.9 to preserve spinodoid morphology in microstructures, as an absolute anisotropy (i.e., $\alpha_e^{mon} = 1$ or $\alpha_{e1}^{ort} = \alpha_{e2}^{ort} = 1$ ) results in truss-like morphologies that deviate significantly from spinodoids.

To solve the volume-constrained compliance minimization TO in Equations (32) and (33), we combine the objective and constraint as the loss function $\ell$ by following the constraint penalty scheme as:

$$\ell(\boldsymbol{\tau}) = J(\boldsymbol{\tau}) + \psi\left[g(\boldsymbol{\tau})\right], \text{ and} \tag{35}$$

$$\psi\left[g(\boldsymbol{\tau})\right] = \left(\eta_0 + n^{opt} \Delta\eta\right) g^2(\boldsymbol{\tau}), \tag{36}$$

where $\eta_0$ and $\Delta\eta$ are user-defined constants ($\eta_0 = 0.05$ and $\Delta\eta = 0.15$), and $n^{opt}$ is the optimization iteration.

We note that classic SIMP and level set based methods may suffer from mesh-dependency issues from their FEA discretization. Mesh dependency could be addressed by using filters (either on the density or sensitivity fields) or using a mesh-free approach [50]. However, it has been shown in [49], [51] that NNs-based TO is able to achieve mesh independence without filters. This is because, different from SIMP which uses element density as design variables, NNs-based TO uses networks' weight parameters as design variables (design reparameterization). One interesting feature is the spectral bias that NNs tend to learn low-frequency (smooth and slowly varying) patterns better than high frequency (rapidly varying) patterns. Therefore, NNs-based TO prefers designs of large feature sizes and appears as mesh-independence, compared to the designs with small features exhibiting checkerboard patterns.

As discussed in Sections 3.1-3.5, our data-driven design framework relies on several important components. We provide a detailed description of the necessary steps to systematically integrate these components to solve spinodoids-based multiscale TOs as shown in Algorithm 1.

**Algorithm 1** Structure of the proposed data-driven multiscale topology optimization for spinodoid architected materials

- Initialization:
  - Generate material data repository containing three types of spinodoids by the DoE and microstructure reconstruction as discussed in Section 3.1, 3.3 and 3.4.
  - Train a GP regression model per spinodoid type from the above material data repository as discussed in Section 3.4.
  - Discretize design domain by a macroscale finite element mesh where each element is associated with a spinodoid microstructure and formulate the design objective and constraints as in Section 3.5.
- Start topology optimization loops:
  - Use NNs to parametrize spinodoid design parameters as NNs' weight parameters, including spinodoid types, material densities, spatial frequencies, anisotropic indices, and rotational angles as in Section 3.5.
  - Use GPs to provide online predictions of effective constitutive constants for spinodoid microstructures with different SDF descriptors as in Section 3.4.
  - Penalize spinodoid elastic properties following the SIMP scheme as in Section 3.5.
  - Perform structure-level FEAs to compute design objectives and constraints and compute the loss by combining objective and constraint values via a constraint penalty scheme as in Section 3.5.
  - Minimize the loss with respect to the NNs parameters via automatic differentiation and update the values of NNs' parameters as discussed in Section 3.5.
  - Repeat the optimization loop until convergence.
- Postprocess:
  - Retrieve spinodoid descriptors and construct bi-continuous spinodoids with continuous microstructure interfaces as discussed in Section 3.2.

## 4. Numerical Experiments

In this section, we present several numerical studies by applying our proposed framework to design multiscale spinodoid architected materials. In Section 4.1, we start to validate our design framework by performing TO on a simple tensile bar where we demonstrate the advantage of multiscale design over single-scale counterparts and quantify the difference of applying different spinodoid types. In Section 4.2, we apply our design framework to a pure bending beam with a detailed discussion on the optimized structural topology and local material morphology. In Section 4.3, we discuss the spinodoids design involving multi-loads boundary conditions and provide physics-consistent design insights. In Section 4.4, we apply our design framework for the design of a hip prosthesis. Compared with classic prostheses, our method provides a novel concept for a porous, customizable, functionally graded design with high stiffness-to-weight ratios.

In our experiments, our date repository is constructed by the spinodoid microstructure reconstruction written in MATLAB [52], and the NNs-based TO is implemented by JAX [53].

We assume that the base material studied in this work has a linear elastic behavior with elastic properties as:

$$E = 1.0\text{e}9 \text{ Pa}, \qquad \nu = 0.35, \tag{37}$$

where $E$ and $\nu$ are Young's modulus and Poisson's ratio of the base material, respectively. All the numerical experiments are carried out on a 64-bit Linux workstation with 18 Intel Xeon W-2295 CPU cores running at 3.0 GHz with 256 GB of RAM.

### 4.1. Design of Tensile Bar

To validate our proposed model, we first apply it to design a tension bar as shown in Figure 15(a). For FEA, the design domain of the tension bar is discretized by $80 \times 20$ quadrilateral elements where each element has four Gaussian integration points. The domain is subject to a fixture boundary condition at its left edge and a horizontal tension force $f = 5e5\ N$ at the center of its right edge.

For the multiscale TO, we assume each of the finite elements of the tensile bar is associated with a spinodoid architected material. We assume the design objective and constraint are the elastic compliance and volume fraction constraint as in Equations (32) and (33).

To compare the performance of the three types of spinodoids, we perform the following experiments. We first carry out a classic single-scale structural TO without considering any architected materials and demonstrate its result in Figure 15(b). In this optimization, we assume the range of the density of macroscale elements as $0 \leq \rho_e^M \leq 1$. But for a fair comparison with spinodoid design whose microscale design $\rho_e^m$ tops at 70% as in Equation (34), the Young's modulus of the single-scale TO is penalized as the value of 70% volume fraction by following the SIMP-based penalization scheme.

From Figure 15(b), it is clear that under the central tension force, solid materials are concentrated along the horizontal center line, consistent with our intuition. We then perform the multiscale TO by using each of the three spinodoid types, namely, isotropic, monoclinic, and orthotropic, as shown in Figure 15(c)-(e), respectively. By observing Figure 15(c1), (d1), and (e1), we find their structural topologies are like the single-scale design in that spinodoids are allocated near the horizontal center line.

What makes the multiscale design different from the single-scale design is that the spinodoid design appears thicker than the single-scale design. This is because compared to the solid materials (with up to 100% macroscale volume fraction) in the single-scale design, the spinodoids are porous materials that can only have up to 70% composed density as in Equation (28). Therefore, more spinodoids are allocated on the outer rims of the tensile bar, therefore increasing its thickness.

We note that, compared to the classic single-scale design, our multiscale spinodoid designs result in spatially varying microstructures, as seen in Figure 15. Another interesting observation is that compared to the isotropic design in Figure 15(c2), both the monoclinic spinodoids in Figure 15(d2) and orthotropic spinodoids in Figure 15(e2) manifest strong direction dependency. Specifically, as the material direction of the monoclinic spinodoids in Figure 15(d2) can only point to one direction, such direction is consistent with the loading path direction (principal stress trajectory [4]) from the external force $f$ as in Figure 15(a).

Comparatively, the material distributions of the orthotropic design can point to two orthogonal directions. Therefore, we can observe from Figure 15(e2) that the materials point to both the loading path direction (horizontally) and a second direction orthogonally (vertically). Additionally, by comparing Figure 15(d2) with Figure 15(e2), we observe that the pores in the orthotropic spinodoids appear shorter and less direction-dictated, resulting in well-connected materials in both horizontal and vertical directions compared to the monoclinic counterpart.

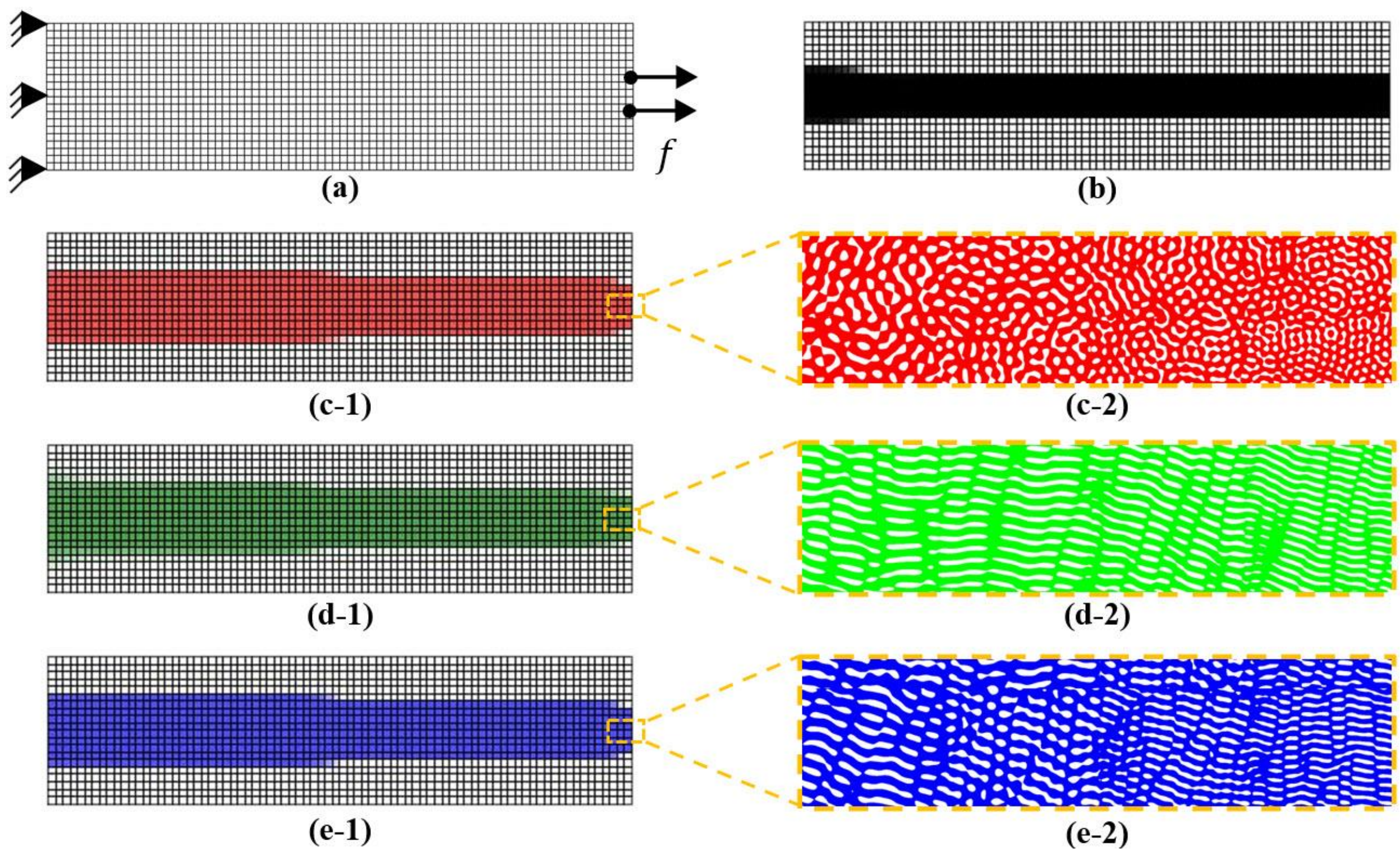

**Figure 15 Design setup and results of the tensible bar experiment: (a)** The design domain is discretized by an $80 \times 20$ finite element mesh where each element is assumed to associate with a spinodoid architected material, and the domain is subject to a fixture boundary condition on the left edge and horizontal tension loads at the center of the right edge; **(b)** The single-scale TO without architectural materials; **(c)** The multiscale design using isotropic spinodoids with structural topology shown in **(c-1)** and local material morphology in **(c-2)**; **(d)** The multiscale design using monoclinic spinodoids; and **(e)** The multiscale design with orthotropic spinodoids.

We compare the convergence histories of the above experiments as in Figure 16. In terms of the design objective, i.e., elastic compliance in Figure 16(a), we observe that all three spinodoids-based multiscale designs result in better compliance values than single-scale design. This is consistent with our expectation, as the design space of multiscale design is larger than that of the single-scale design. Multiscale designs, therefore, may end up with better optimum solutions.

When comparing the compliance among the three different spinodoid types, we note that the compliance of the monoclinic design ($J^{mon} = 4.85e3\ \mathrm{N \cdot m}$) is significantly better (lower) than the others, while the orthotropic spinodoids ($J^{ort} = 7.67e3\ \mathrm{N \cdot m}$) is slightly better than that of the isotropic counterpart ($J^{iso} = 7.86e3\ \mathrm{N \cdot m}$). This comparison is meaningful. Because it indicates that the monoclinic spinodoids have the strongest unidirectional dependency, as their morphology is distributed along the (unidirectional) principal stress trajectory.

We proceed to perform an experiment to let the optimizer freely choose the spinodoid type. We find the final design is identical to the monoclinic design as in Figure 15(d). Therefore, we find monoclinic spinodoids are the optimal architected material candidates, when structural loading paths exhibit mostly uniform directions.

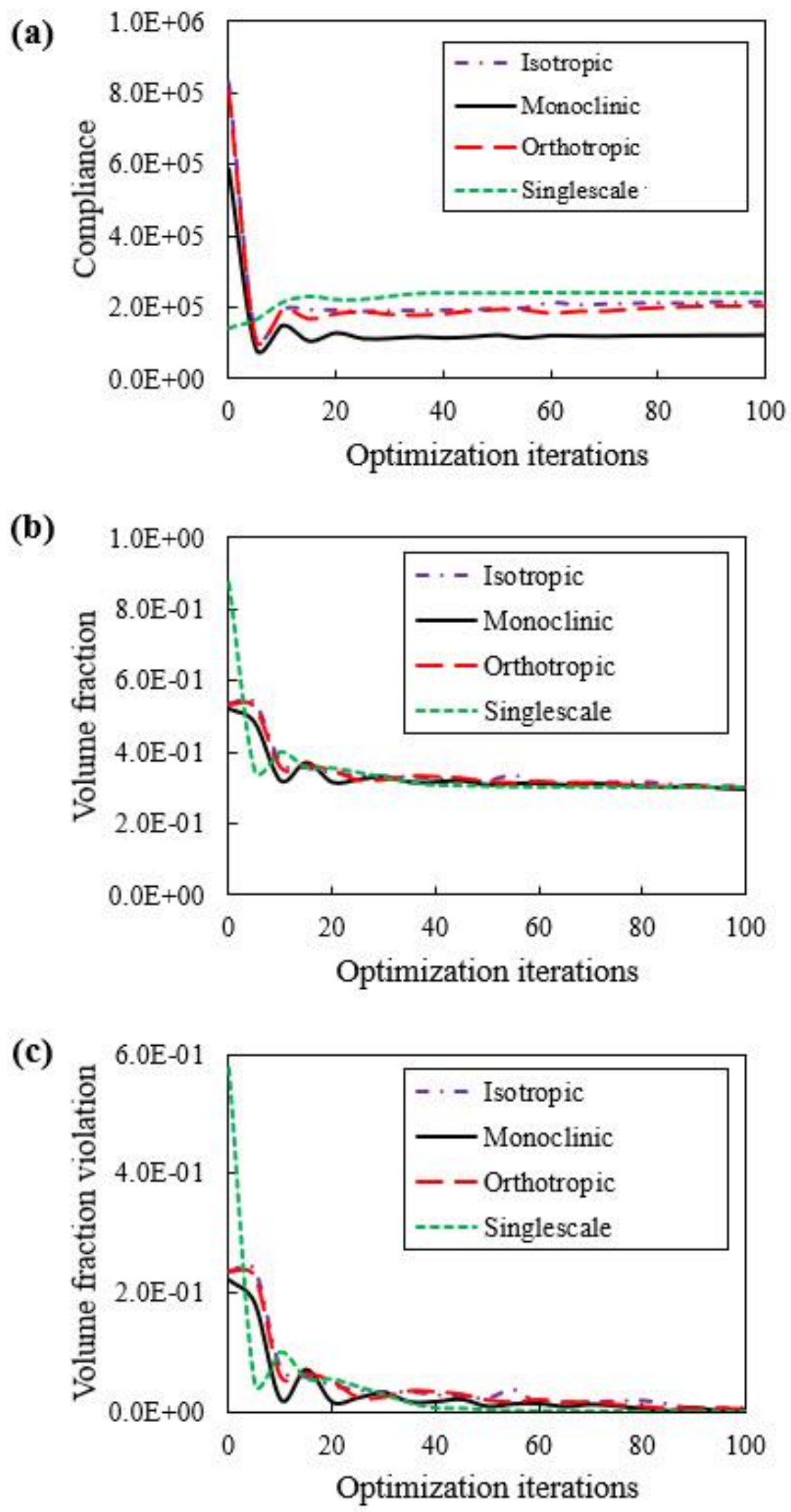

**Figure 16 Convergence of the tensible bar experiments: (a)** Comparison of the optimization histories for the objective values of elastic compliance for the four architected material settings, **(b)** Convergence histories for the volume fraction constraint, and (c) Convergence histories of the absolute value of volume fraction violations.

On the contrary, the orthotropic and isotropic spinodoids show much weaker directional dependencies, as their materials generally point to more than one direction. Thus, the orthotropic spinodoids are preferred at locations exhibiting two orthogonal loading paths, while the isotropic architected materials are preferred in regions with multiple different principal stress trajectory directions. We use the experiments in Sections 4.2 and 4.3 to demonstrate the usage of the three different types of spinodoid architected materials.

From Figure 16(b), our penalty-based optimization scheme in Equation (35) is able to enforce the volume fraction constraint as the optimization proceeds. It is noted that although such a scheme is proven to work for our current single-objective and single-constraint setup, more advanced optimization techniques are preferred, e.g., the augmented Lagrangian method [54], [55], [56], sequential quadratic programming [41], method of moving asymptotes [57] or Newton-based methods [41], for more challenging optimization problems, involving multiple constraints, multiple objectives, objective sign changing, large scale and non-convexity.

We note that we used a penalty scheme to solve the (volume fraction) constrained (loss) minimization problem. We demonstrate the convergence history of the absolute value of the volume fraction violations in Figure 16(c). In our work, the volume fraction constraint was treated as a soft constraint. That is, there is no guarantee that the constraint is always satisfied during optimization iterations. Once violated, the constraint will penalize the loss by assigning higher weights to the constraint term in the following optimization iterations.

We also want to note that our NNs-based TO may experience slow convergence. This is because our NNs-based TO tends to learn low-frequency designs better than high-frequency ones. Therefore, if there are fine design details with sharp features, it may take many optimization iterations to capture high-frequency design features. An alternative approach to addressing slow convergence issue could be found at [58].

### 4.2. Design of Bending Beam

In this experiment, we perform multiscale TOs for a pure bending beam to demonstrate the usage of different types of spinodoid architected materials. We assume the design domain of the beam is discretized by a $40 \times 20$ mesh of quadrilateral fully integrated finite elements, as in Figure 17(a). The beam is fixed at the left edge and a bending moment $M = 1e6\ \mathrm{N \cdot m}$ is applied to the center of the right edge.

Similar to the previous experiment in Section 4.1, the objective and constraint of this optimization are the elastic compliance and volume fraction, respectively. We assume that each of the macroscale finite elements on the beam structure in Figure 17(a) is represented by a spinodoid architected material. Different from the previous experiment, we do not limit the spinodoid types and let the optimization freely choose from the three types, namely, isotropic, monoclinic, and orthotropic spinodoids, for each macroscale finite element.

The distributions of elemental densities, deformed structural topology, spinodoids types, and local material morphology are illustrated in Figure 17(b) and Figure 18(a), respectively. From Figure 17(b), we find the final design of the bending beam is similar to the shape of the I-beam. In this design, materials are optimized to allocate at the top and bottom rims of the design domain which is subject to larger normal stresses compared to internal areas closer to the neutral axis. Additionally, if we take a closer look at the top or bottom rims, the elements close to the outer surface manifest even higher material densities than the interior ones, as the outer surface is subject to the highest tension and compression stresses from the bending moment. Furthermore, we note that the densities on the right edge are smaller than those of the top and bottom rims. This is because the magnitude of normal stresses close to the neutral axis are much smaller than top and bottom regions.

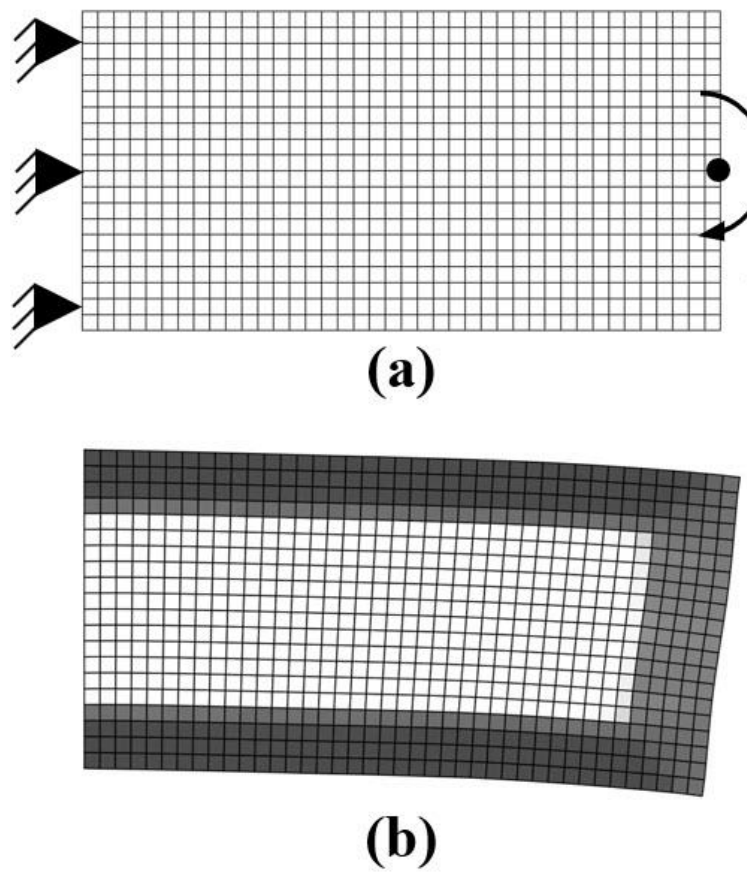

**Figure 17 Bending beam experiment: (a)** The design domain is discretized by a $40 \times 20$ finite element mesh which is fixed at the left edge and subject to a bending moment at the center of the right edge, and **(b)** The density distribution after TO where each finite element can choose any spinodoid types.

On the one hand, in Figure 18(a), we find monoclinic spinodoids (color coded as green) are distributed at the top and bottom rim regions. This observation is consistent with our expectation, as those regions are mostly subject to unidirectional tension and compression normal forces (with the minimum of shear forces). As we discussed in Section 4.1, monoclinic spinodoids with strong directional anisotropy are the best spinodoids candidate to withstand single-directional loading.

On the other hand, isotropic spinodoids (color coded as red) are found at the center of the right edge where the bending moment is applied. This is because in regions close to the bending moment, there are multiple different loading path directions, arising from the combinatorial effects of (tension and compression) normal stresses, and high shear stresses near the neutral axis.

The material morphology in Figure 18(a-2) provides a good indication of the local principal stress trajectory. For example, the monoclinic spinodoids on the top part of Figure 18(a-2) show a loading path pointing diagonally downward. This is because it is subject to a horizontal tension normal force and a downward shear force simultaneously. The material morphology of the isotropic spinodoids close to the neutral axis indicates multiple local loading path directions, as shear stress tends to reach the maximum and normal stress reduces to zero. We also illustrate continuous transitions between microstructure interfaces in Figure 18(a-3), demonstrating the importance of applying the interface interpolation and amplifying functions as discussed in Section 3.2.

For comparison, we perform the same optimization experiment, but only using monoclinic spinodoids as shown in Figure 18(b). While the structural topology of Figure 18(b-1) remains unchanged, the local material morphology in Figure 18(b-2) and (b-3) shows significant difference. Different from Figure 18(a-2) which separates a strong unidirectional monoclinic region and an isotropic region, the monoclinic spinodoids of Figure 18(b-2) show a much weaker unidirectional dependency, as the monoclinic spinodoids have to simultaneously account for loading paths along different directions. It is very interesting to note that in Figure 18(b-3), local materials on the neutral axis mostly points to the vertical direction, as (vertical) shear stress dominates (horizontal) normal stress.

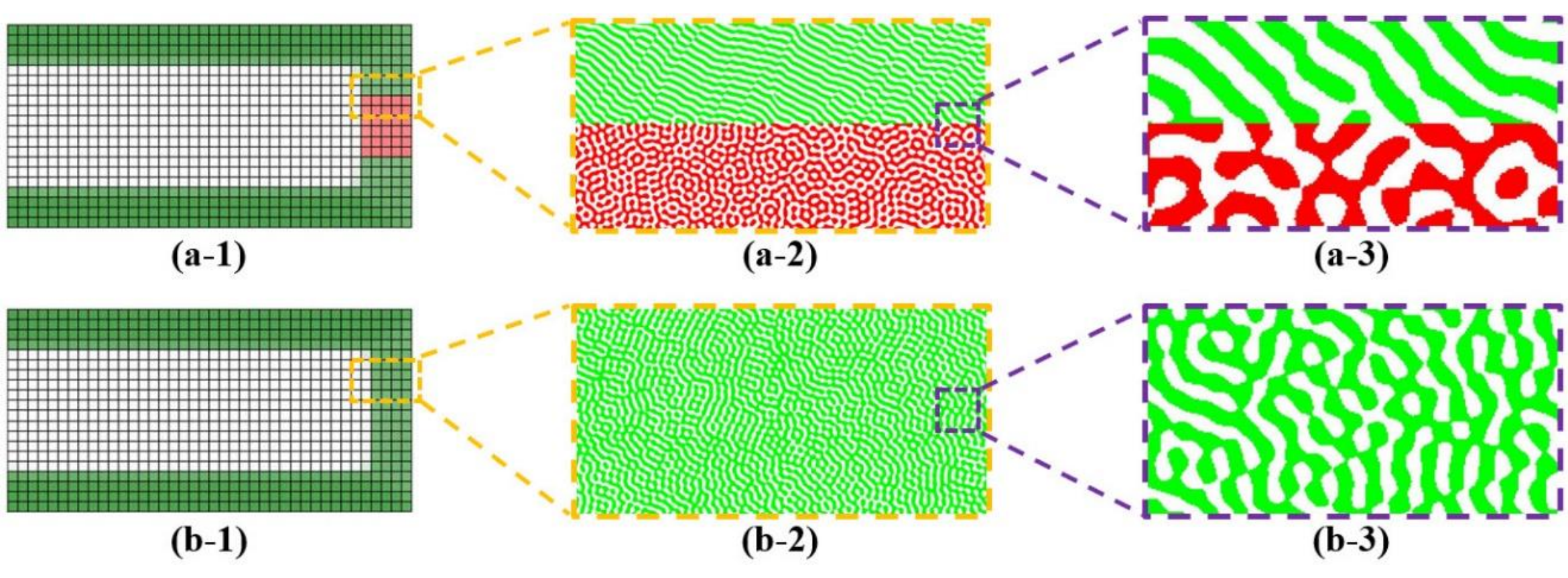

**Figure 18 Multiscale design results of the beam bending experiment: (a)** Design using multiple types of spinodoid materials where monoclinic spinodoid materials are optimized to locate on the top and bottom sections with the higher stress magnitude and the isotropic architected is located around the bending location with lower stress concentrations as shown in **(a-2)**, and smooth boundary interfaces after applying interpolation and amplifying functions are illustrated in **(a-3)**; **(b)** Design optimization using only monoclinic spinodoids with structural topology design in **(b-1)**, local morphology design in **(b-2),** and boundary interface in **(b-3)**.

We proceed to compare the convergence histories of the objective compliance for the two models of Figure 18(a) and (b), as shown in Figure 19. We find that the compliance of the multiscale model using multiple types of spinodoids is slightly better than that of monoclinic spinodoids. A plausible reason why monoclinic spinodoids perform only slightly worse could be because: both models result in strong unidirectional dependent monoclinic spinodoids in the top and bottom rim regions that can withstand the largest magnitude of normal tension and compression forces, while the morphology difference (isotropic versus monoclinic) near the neutral axis only contribute to a slight difference of the overall compliance.

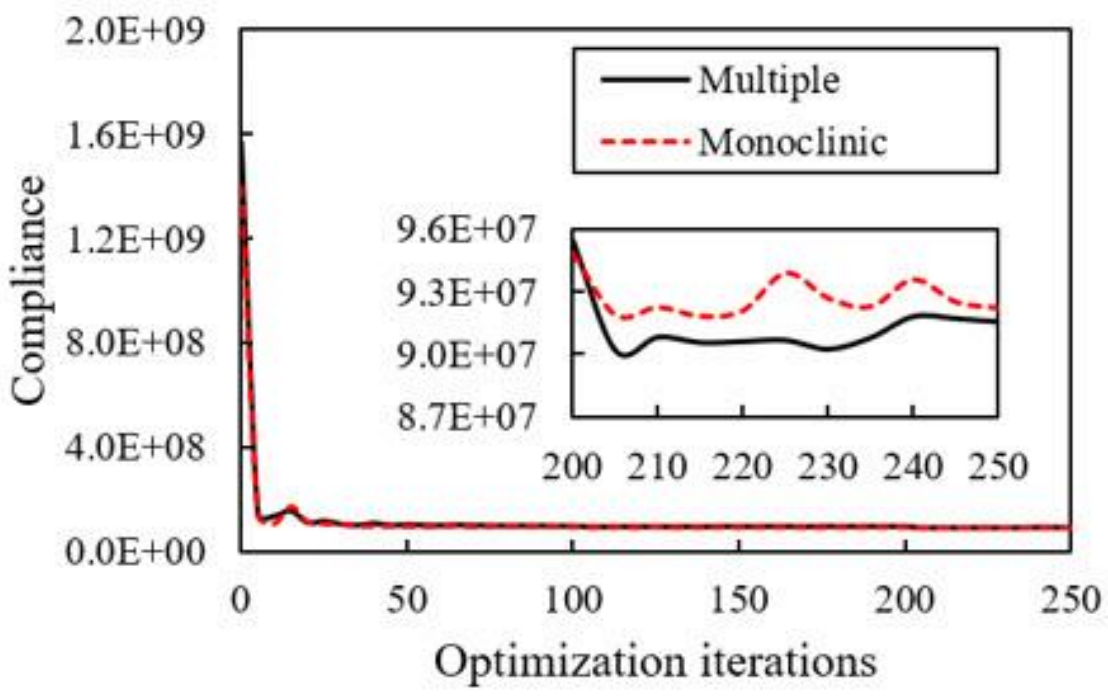

**Figure 19 Convergence of the beam bending experiment:** Compared with the monoclinic material design, the design using both isotropic and monoclinic materials shows slightly better compliance value.

### 4.3. Design under Multi-Loads

In this experiment, we aim to demonstrate the multiscale TO of different types of spinodoids when the material-structure system is subject to multi-loads boundary conditions. The design domain is illustrated in Figure 20(a) which is discretized by a $40 \times 40$ mesh. Similar to previous experiments, each macroscale element in Figure 20(a) has four Gaussian integration points, and it is assumed to be associated with a spinodoid architected material whose type can be freely chosen from isotropic, monoclinic, and orthotropic types. The domain is fully fixed at its left edge, and two loading cases (a tension force $f_1$ and a bending force $f_2$) are applied separately to the center of the right edge.

Similar to the previous examples, a 30% material volume fraction constraint is enforced in the optimization. We note that this is a multi-load scenario where the objective compliance is taken as the sum of the individual compliance associated with each of the two loads. That is, it is not a multiple-loads scenario where multiple forces are applied simultaneously which has only one resulting displacement field (see Section 4.4 for an example).

To compare the optimization results, we perform two experiments using different loading settings. In the first experiment in Figure 20(a), we let the tension force dominate the bending force as $f_1 = 50e5$ N and $f_2 = 1e5$ N. The resulting topology after optimization is shown in Figure 20(b), which is very similar to the tension bar experiment in Section 4.1. Similarly, all macroscale elements choose monoclinic spinodoids, as they exhibit strong unidirectional anisotropy. Different from the pure tension example, this model is also subject to a vertical force, which results in a bullet shape topology.

In the second experiment, we let the bending force be much larger than the tension force by setting $f_1 = 0.5e5$ N and $f_2 = 1e5$ N. This loading condition results in a triangle-shaped structural design as in Figure 20(c). Interestingly, the monoclinic spinodoids occupy the upper arm and lower arm regions, while orthotropic spinodoids are located along the neutral axis in the central region.

We plot local microstructures in Figure 20(d)-(f) to demonstrate local material morphology. The spinodoids in Figure 20(d) is at the bottom left corner of the structure. When we take a close look at its material orientations, we find that its morphology mostly points from the fixture boundary (left edge) to the force location. More interestingly, the bottom half of the spinodoids (close to surfaces) points horizontally to the right, parallel to the bottom surface of the structure, while the top half (far away from surfaces) mostly follows loading path directions pointing to the force-applying location. A similar observation can be found in Figure 20(e) where local monoclinic spinodoids are far from any surfaces and mostly point to the force-applying location. In other words, TO controls the directional anisotropy in those monoclinic microstructures.

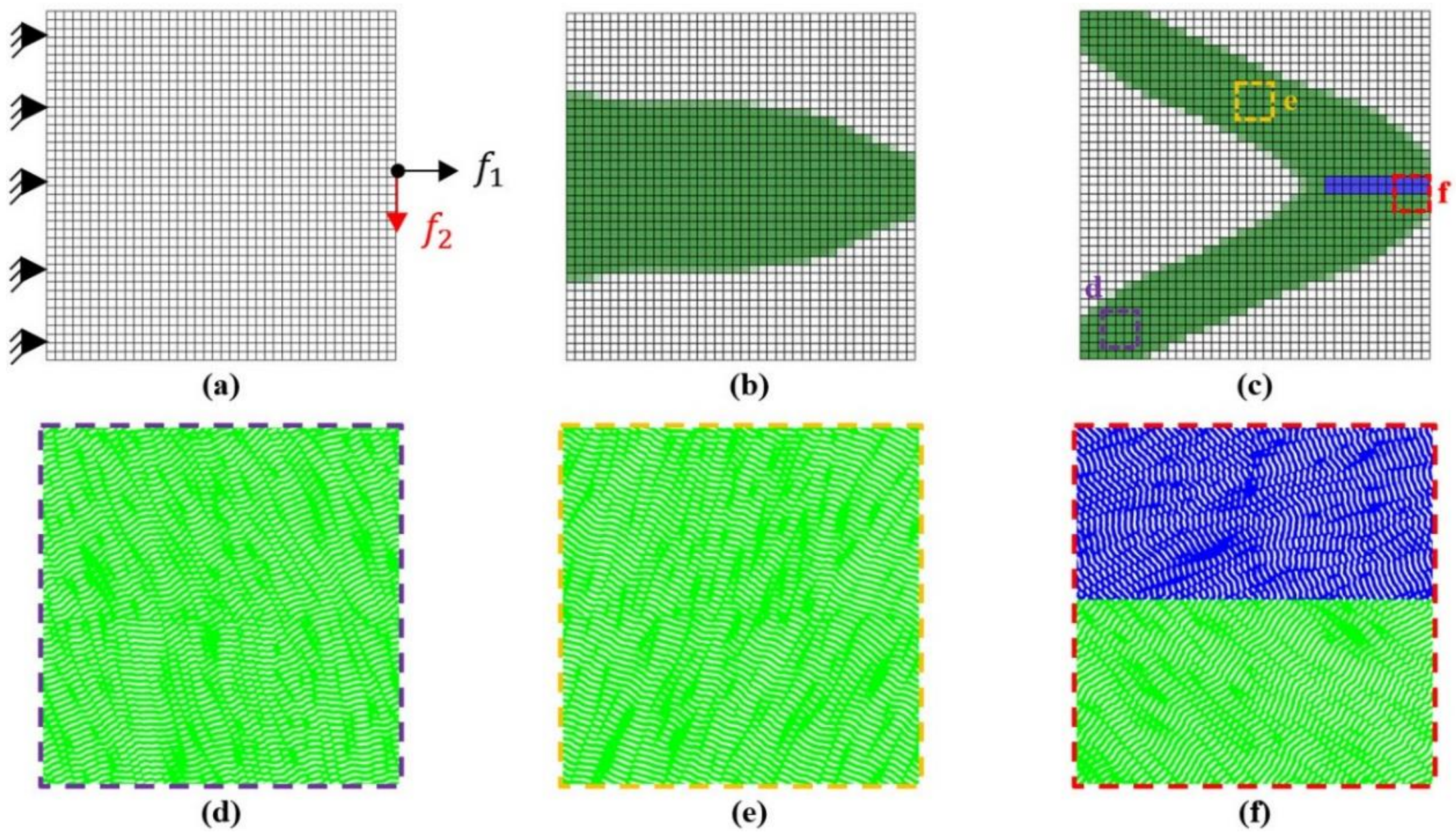

**Figure 20 Design setup and design results of the multi-loads experiment: (a)** The design domain is discretized by a $40 \times 40$ finite element mesh where each element is assumed to be associated with a spinodoid microstructure. While the domain is fixed at the left edge, the center of the right edge is subject to two loading cases: a horizontal force $f_1$ and a vertical force $f_2$; **(b)** When the magnitude of $f_1$ is much larger than $f_2$, TO results in a design similar to tensile bar design where all spinodoids have the monoclinic type, as we discussed in Section 4.1; and **(c)** When the magnitude of $f_1$ is much smaller than $f_2$, the optimization results in a triangle-like design where monoclinic spinodoids are found in the top and bottom arm regions, and the orthotropic spinodoids are found in the middle region where two loads are separately applied. The local material morphologies are illustrated at three different locations as shown in **(d)**, **(e),** and **(f)** with clear TO-controlled directional anisotropy.

In Figure 20(f), the spinodoids exhibit different types in the top and bottom parts. At the bottom part, it is the monoclinic spinodoids that follow the principal stress trajectory generated by the compression normal stress and vertical downward shear stress from the bending force $f_2$, as we discussed in Section 4.2. At the top part close to the neutral axis, TO chooses the orthotropic spinodoids (color-coded as blue) that generally point to horizontal and vertical directions. We note that this loading scenario is different from the pure bending case in Section 4.2. In Section 4.2, the pure bending results in multiple normal and shear forces which can be considered as a multiple-loads scenario, and the loading path direction is determined by the total force of the normal and shear forces. Here, the tension force $f_1$ and bending force $f_2$ are applied in a multi-loads scenario such that local materials must interpedently account for the two orthogonally oriented forces. That is the reason TO chooses the orthotropic spinodoids over isotropic counterparts.

We then illustrate the evolving history of the spinodoid types in Figure 21. We observe that while the spinodoid types are randomly initiated in the beginning, orthotropic spinodoids gradually cluster along the neutral axis close to the force-applying location, while monoclinic spinodoids takeover other regions to withstand local loads with mostly unidirectional principal stress trajectories.

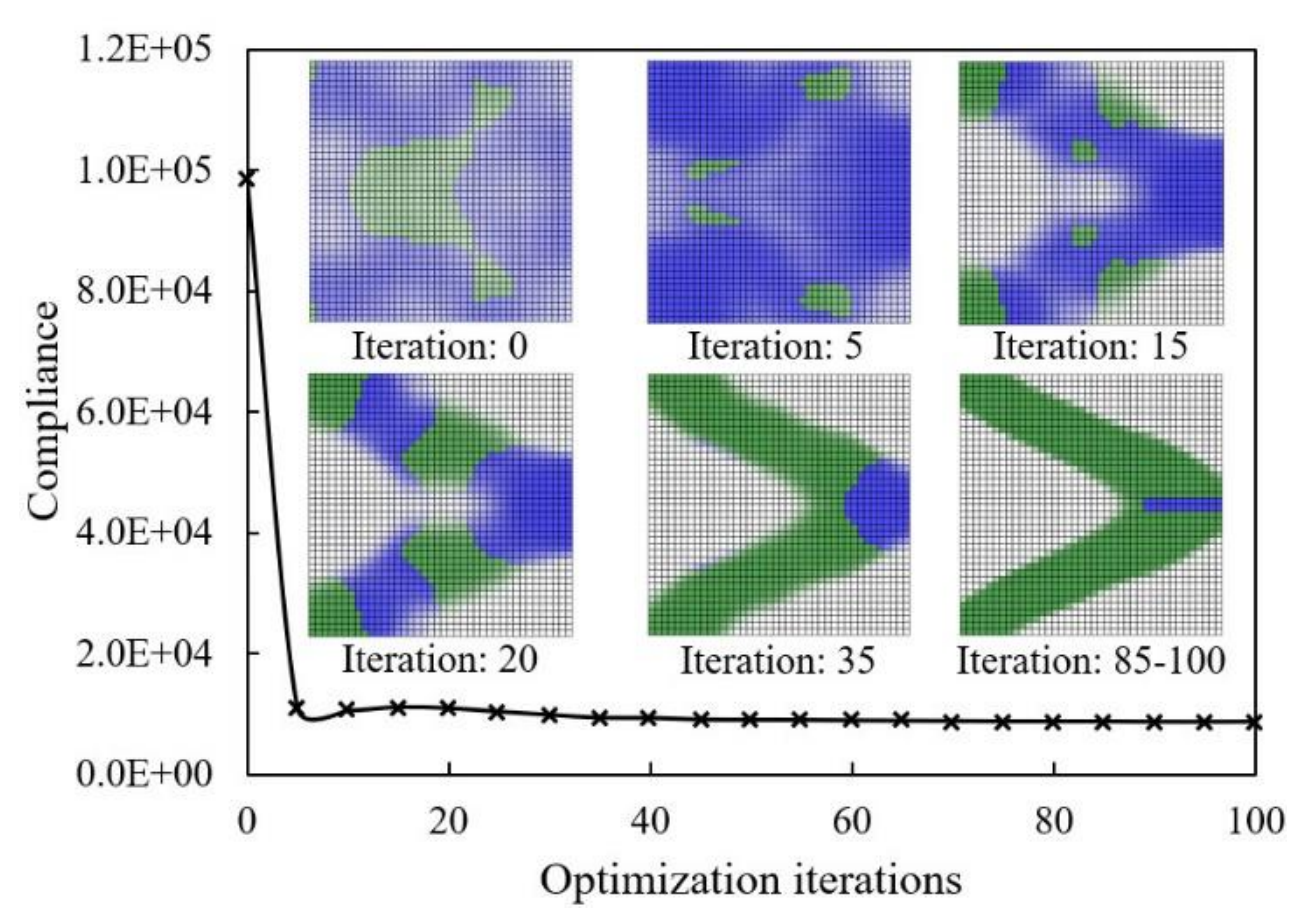

**Figure 21 Convergence of the multi-loads experiment:** The evolving history of the distributions of spinodoid types on the design domain is demonstrated at different optimization iterations.

### 4.4. Design of Hip Prosthesis

To further demonstrate our framework, we apply it for the design of femur prostheses in this section. Traditional femur prostheses are often designed as solid bulk structures made of metallic alloys [59], [60]. While metallic alloys are durable, corrosion resistant, and bio-compatible, they are generally heavy and lacking customization. Additionally, as alloys are much stiffer than bones, stiffness mismatch may cause bones to experience reduced loading and result in bone resorption. To improve the prosthesis design, we use our proposed multiscale TO framework to design a lightweight prosthesis structure with spatially varying spinodoid microstructures.

The model of the femur prostheses is illustrated in Figure 22(a). Its design domain is discretized by 2,032 quadrilateral finite elements, and each element is assumed to be associated with one spinodoid microstructure. As shown in Figure 22(b), the design domain is subject to multiple loads, including two side-way compression forces $f_1 = 1e5$ N and $f_2 = 1e5$ N, and a shear force $f_3 = 1e5$ N on the bottom surface.

Like the previous numerical examples, our design objective is the elastic compliance, and the design constraint is the 30% material volume fraction. Our optimized structural topology is demonstrated in Figure 22(c). In this multiple-loads experiment, we find all final spinodoids belong to the monoclinic type.

To demonstrate directional anisotropies, we plot out the anisotropic rotational angles (see $\gamma$ in Equation (17)) of each monoclinic spinodoid microstructure as in Figure 22(d) and an enlarged image in Figure 22(e). It is observed that the directional anisotropic angles (shown as double headed arrows) are spatially varying across the structure, as they point along the directions of principal stress trajectories at different locations.

We also demonstrate material morphologies at three locations of interests, as shown in Figure 22(f)-(g), respectively. Firstly, microstructures at Figure 22(f) are located on a truss-like region connecting the fixture boundary at the top and the compression load $f_1$. They manifest strong directional anisotropies, as their monoclinic rotational angles point along the direction of the truss-like region. Secondly, the microstructures at Figure 22(g) is located on a pathway between the compression load $f_2$ and the shear force $f_3$ at the bottom. We can observe a clear directional dependency of their morphologies along the vertical direction of the pathway. But their anisotropies (see $\alpha^{\text{mon}}$ from Equation (6)) appear weaker than that of Figure 22(f). This is because spinodoid microstructures of Figure 22(g) are in proximity of the compression force $f_2$, which causes local materials to exhibit a principal stress trajectory pointing to the location of $f_2$. Thirdly, microstructures of Figure 22(h) are at a joint location connecting two truss-like pathways from the top and close to the shear load $f_3$ at the bottom. We can see the anisotropy of those microstructures tend to simultaneously account for the directions of principal stress trajectories from Figure 22(g) and Figure 22 (h). We note that since the directions of the two principal stress trajectories are not orthogonal, TO selects monoclinic spinodoids with small anisotropic indices (see $\alpha^{\text{mon}}$ from Equation (6)) than orthotropic spinodoids with two orthogonal anisotropic directions.

In this section, our multiscale TO provides a new design concept as in Figure 22(c)-(h) for a porous design with functionally graded spinodoids. Compared to the classic prosthesis design in Figure 22(a), our design is lightweight, customizable for different design domains and loading conditions per patient. In addition, its spatially varying spinodoid microstructures can achieve functionally graded designs which may mitigate stiffness mismatch between the prosthesis and bone, thereby reducing the risk of bone resorption.

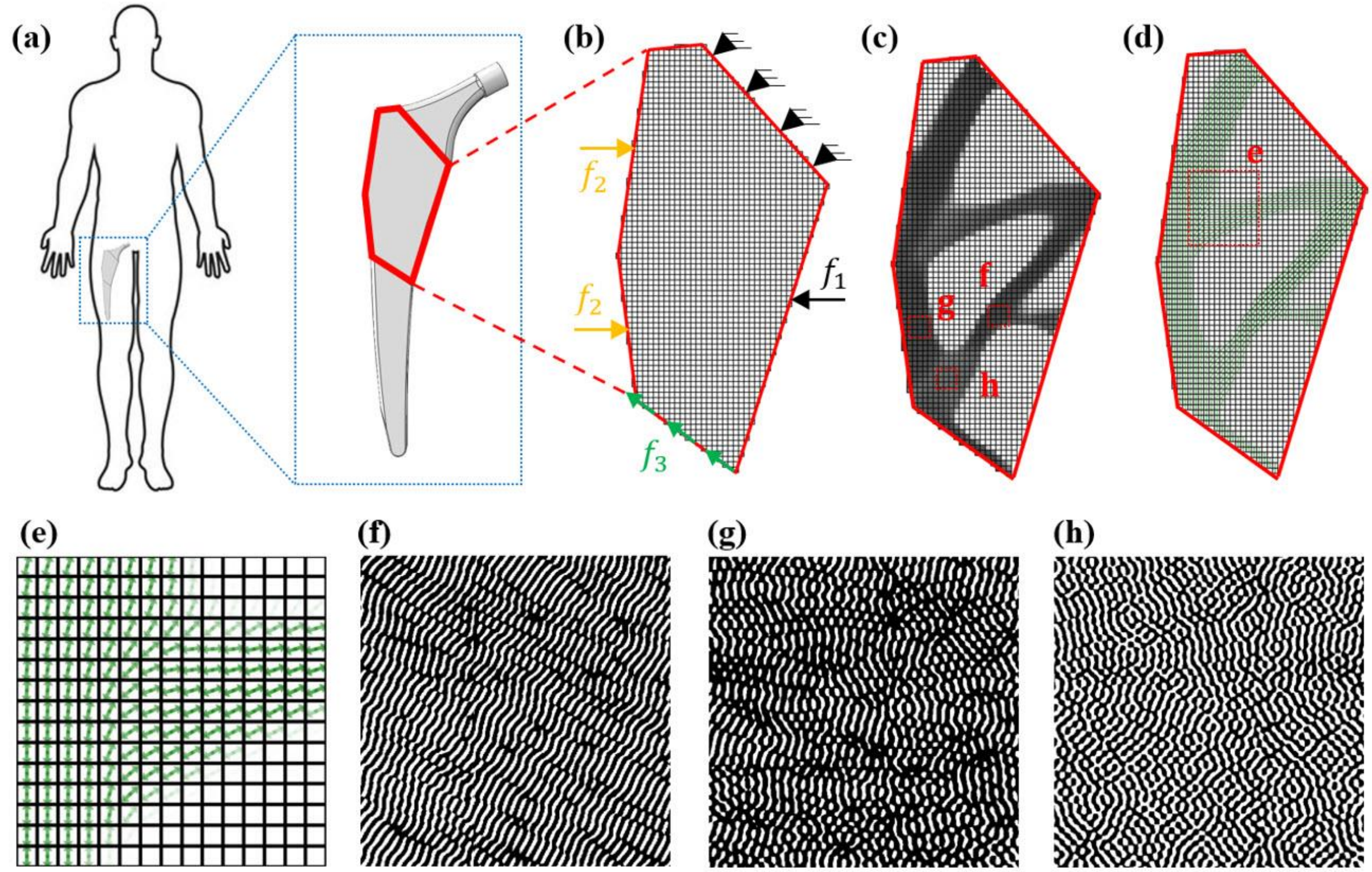

**Figure 22 Design setup and design results of the prosthesis: (a)** Design domain; **(b)** Discretized mesh and boundary conditions; **(c)** Optimized structural topology; **(d)** Streamlines representing anisotropic orientational angles; **(e)** Streamlines at an enlarged region; and **(f-g)** Local material morphologies at the locations of interests.

## 5. Conclusion

In this work, we propose a data-driven design framework for 2D spinodoid architected material systems with controllable anisotropic properties. Specifically, we develop an NNs-based multiscale TO that enables automatic differentiation to streamline the computation of design gradients with respect to various spinodoid descriptors. Additionally, we introduce a low-dimensional SDF-based microstructure reconstruction method, combined with an interface interpolation scheme, to efficiently generate three types of spinodoids with aperiodic, stochastic, bi-continuous, and directional anisotropic properties. We also develop GP-based data-driven surrogates of spinodoid constitutive models to bypass repetitive computational homogenization in our multiscale TO framework.

A primary strength of our approach is to reduce the cost of computational homogenization when considering spinodoids as unit cells in multiscale TOs. To circumvent expensive real-time homogenization, we develop GP-based surrogate models, which predict spinodoids' effective properties as functions of their SDF descriptors. Our approach is similar to the pioneering work in [4], which utilized polynomial surrogates to link local material volume fraction to effective properties. Compared to direct numerical simulations (e.g., $FE^2$ methods), constitutive surrogate models often reduce computational costs by up to four orders of magnitude [18], [61], [62], enabling rapid design optimization.

Through several numerical examples, we demonstrate that spinodoid designs with tailored directional anisotropy outperform single-scale and isotropic alternatives while providing clear, physics-consistent design insights. Unlike 'black-box' deep learning approaches, our method provides clear physical insights into material distribution: it explicitly reveals why anisotropic spinodoids with tailored orientations are favored in specific regions, while isotropic spinodoids are optimally placed elsewhere. This interpretability bridges the gap between data-driven design with mechanistic understanding. Additionally, our experiments demonstrate that our model is applicable to a wide range of biomimetic designs, utilizing porous, functionally graded spinodoid architected materials with high stiffness-to-weight ratios.

To compare our framework with existing approaches, we will consider improvement of both performance and computational efficiency. In terms of the performance improvements, it would be ideal to compare the objective values of the optimized spinodoid structures with truss-, beam- and plate-based mechanical metamaterials.

Theoretically, the performance of our spinodoids should be no worse than such metamaterial designs. This is because our spinodoids offer free-form designs without predefined topologies, in contrast to traditional metamaterials, which typically utilize parameterized geometries with tunable dimensions (e.g., thickness of plates and cross-sectional areas of beams). A comprehensive comparison will be considered in our future work considering changing parameters of a variety of geometrical configurations.

We would like to point out that one of the key distinctions between our proposed spinodoids and traditional truss-based metamaterials lies in the incorporation of local morphological randomness. This randomness is not incidental - it is a deliberate design feature that enhances robustness to manufacturing imperfections, making spinodoids potentially more resilient in real-world applications.

While it is indeed meaningful to quantitatively compare how process-induced randomness affects microscale properties in spinodoids versus truss-based metamaterials, such a comparison would require a dedicated and systematic study. Specifically, it would involve uncertainty quantification and propagation across multiple stages, i.e., from process parameters to structural features, and ultimately to material properties and performance. We view this as an important direction for future work.

Moreover, we acknowledge that in terms of linear elastic performance, spinodoids may not outperform classic truss- or beam-based metamaterials. The latter are often more efficient in aligning material along principal stress paths, whereas spinodoids, by design, introduce additional material to achieve randomness, which may not always align with loading directions. Therefore, a direct comparison of elastic properties alone may not fully capture the design intent or advantages of spinodoids.

For these reasons, we believe that a simple quantitative comparison of elastic behavior would be insufficient and potentially misleading. Instead, we plan to pursue a more comprehensive evaluation using uncertainty quantification methods in future studies to better understand and compare the performance of these architected material systems.

For more future work, we are interested in extending the current framework to: ($i$) study the optimal design of spinodoids under dynamic loading scenarios, including wave guidance and impact resistance; ($ii$) design architected materials in highly nonlinear regimes considering path-dependent constitutive models, such as plasticity [63], damage, and large deformations [64]; ($iii$) build an open-source 2D/3D spinodoids data repository with rich morphology variants and multiple constitutive behaviors; ($iv$) statistically quantify the impacts of manufacturing imperfection [65], [66] on the damage behaviors of spinodoids; ($v$) systematically understand the influence of NNs' architecture and hyper-parameters on the neural TO; and ($vi$) study multi-physics design, e.g., thermal-elasticity [67], [68] and fluid-mass transportation [69], and multi-constrained design [54], [55], [56], e.g., multiple mass and volume fraction constraints for advanced multi-material manufacturing, and multiple displacement constraints in compliant mechanism.

## 6. Statements and Declarations

### 6.1. Acknowledgments

The authors would also like to thank the constructive suggestions from anonymous reviewers.

### 6.2. Funding

Shiguang Deng would like to thank the support of the Faculty Startup Grant and the New Faculty Research Development Grant from the University of Kansas. Liwei Wang acknowledges support from the Department of Mechanical Engineering at Carnegie Mellon University. The Northwestern team would like to acknowledge the acknowledge the support from the DARPA METALS program project RADICAL (HR0011-24-2-0302) and the NSF BRITE fellow grant (2227641).

### 6.3. Replication of results

The results presented should be reproducible through the implementation of the methodologies described in this publication.

### 6.4. Ethics approval and Consent to participate

Not applicable.

### 6.5. Author contributions

Shiguang Deng contributed to visualization, methodology, implementation, validation, writing-original draft, review and editing. Doksoo Lee contributed to methodology, writing-review & editing. Aaditya Chandrasekhar contributed to methodology. Stefan Knapik contributed to methodology. Liwei Wang contributed to methodology, writing-review & editing. Horacio D. Espinosa contributed to visualization, methodology, writing-review & editing, and supervision. Wei Chen contributed to visualization, methodology, writing-review & editing, and supervision.

### 6.6. Data availability

All data involved are included in the paper.

### 6.7. Conflict of interest

The authors declare no competing financial or non-financial interests.